\pdfoutput=1
\documentclass[aps,prd,onecolumn,superscriptaddress,preprintnumbers,nofootinbib, a4paper]{revtex4}
\usepackage[all]{xy}
\usepackage{graphicx}
\usepackage{slashed}
\usepackage{amsmath,bbm,latexsym,amssymb}
\usepackage{epsfig}
\usepackage{epstopdf}
\usepackage{float} 
\usepackage{tensor}
\usepackage{booktabs}
\usepackage{multirow}
\usepackage{cancel}
\usepackage{parskip}

\makeatletter
\renewcommand*\env@matrix[1][\arraystretch]{%
  \edef\arraystretch{#1}%
  \hskip -\arraycolsep
  \let\@ifnextchar\new@ifnextchar
  \array{*\c@MaxMatrixCols c}}
\makeatother

\usepackage{tabularx}
\newcolumntype{C}[1]{>{\centering\arraybackslash}p{#1}}

\newcommand{\Z}[1]{\ensuremath{\mathbbm{Z}_{#1}}} 
\def\beq{\begin{equation}}
\def\eeq{\end{equation}}
\renewcommand{\Re}{\textrm{Re}}
\renewcommand{\Im}{\textrm{Im}}

\usepackage{xparse}
\usepackage{xcolor}
\usepackage[hyperfootnotes=false]{hyperref}
\hypersetup{
    colorlinks=true,
    linkcolor=blue,
    filecolor=blue,      
    urlcolor=blue,
    citecolor=blue
}
\usepackage{bbold}
\usepackage{physics}
\newcommand{\one}{\mathbb{1}}

\begin{document}
\preprint{IPPP/25/28}  

\title{Unearthing large pseudoscalar Yukawa couplings with Machine Learning} 

\author{Fernando Abreu de Souza}
\email{abreurocha@lip.pt}
\affiliation{LIP -- Laborat\'orio de Instrumenta\c{c}\~ao e F\'isica Experimental de Part\'iculas, Escola de Ciências, Campus de Gualtar, Universidade do Minho, 4701-057 Braga, Portugal}
\author{Rafael Boto}%
\email{rafael.boto@tecnico.ulisboa.pt}
\affiliation{Departamento de
	F\'isica and CFTP,  Instituto Superior T\'ecnico,\\  \small\em
	Universidade de Lisboa, Av
	Rovisco Pais, 1, P-1049-001 Lisboa, Portugal}
\author{Miguel Crispim Romão}
\email{miguel.romao@durham.ac.uk}
\affiliation{Institute for Particle Physics Phenomenology, Durham University, Durham DH1 3LE, UK}
\affiliation{LIP -- Laborat\'orio de Instrumenta\c{c}\~ao e F\'isica Experimental de Part\'iculas, Escola de Ciências, Campus de Gualtar, Universidade do Minho, 4701-057 Braga, Portugal}
\author{Pedro N. Figueiredo}%
\email{pedro.m.figueiredo@tecnico.ulisboa.pt}
\affiliation{Departamento de
	F\'isica and CFTP,  Instituto Superior T\'ecnico,\\  \small\em
	Universidade de Lisboa, Av
	Rovisco Pais, 1, P-1049-001 Lisboa, Portugal}
\author{Jorge C. Romão}%
\email{jorge.romao@tecnico.ulisboa.pt}
\affiliation{Departamento de
	F\'isica and CFTP,  Instituto Superior T\'ecnico,\\  \small\em
	Universidade de Lisboa, Av
	Rovisco Pais, 1, P-1049-001 Lisboa, Portugal}
\author{Jo\~ao P.\ Silva}%
\email{jpsilva@cftp.ist.utl.pt}
\affiliation{Departamento de
	F\'isica and CFTP,  Instituto Superior T\'ecnico,\\  \small\em
	Universidade de Lisboa, Av
	Rovisco Pais, 1, P-1049-001 Lisboa, Portugal}

\begin{abstract}
With the Large Hadron Collider's Run 3 in progress, the $125\textrm{GeV}$ Higgs boson couplings are
being examined in greater
detail,
while searching for additional scalars.
Multi-Higgs frameworks allow Higgs couplings to significantly deviate from
Standard Model values, enabling indirect probes of extra scalars. We
consider the possibility of large pseudoscalar Yukawa couplings in the
softly-broken $\Z2\times\Z2'$ three-Higgs doublet model with
CP violating coefficients. To explore the parameter space of the
model, we employ a Machine Learning algorithm that significantly enhances
sampling efficiency. Using it, we find new regions of parameter space and observable consequences, not found with previous techniques. This method leverages an Evolutionary Strategy to
quickly converge
towards valid regions with an
additional Novelty Reward mechanism. We use this model as a prototype to illustrate the potential of the new techniques,
applicable to any Physics Beyond the Standard Model scenario. 
\end{abstract}

\maketitle

\section{Introduction}

The ATLAS \cite{ATLAS:2012yve} and
CMS \cite{CMS:2012qbp} Collaborations at the Large Hadron Collider (LHC)
had their first success with the discovery of the Higgs boson at $125 \textrm{GeV}$
($h_{125}$).
This enticed two new questions:
i) are the properties of this scalar consistent with those predicted by the Standard Model (SM)?;
ii) are there more scalar families, just as there are extra fermion families?
The first question calls for precision experiments, the second for exploratory searches.
And one question informs the other;
precision experiments probe the effect of extra particle through their putative
virtual effects;
models with extra scalars inform exploration for unusual coupling properties.

Both questions are actively being pursued,
and considerable new knowledge has been acquired from the LHC.
One has learned that the magnitudes of the tree-level couplings of the $125 \textrm{GeV}$ Higgs
to all members of the third fermion family and to a pair of weak gauge boson is
in accordance with the SM, to a precision of about $10 \%$.
Loop couplings with a pair of gluons and with a pair of photons have also
been probed to considerable precision \cite{ATLAS:2022vkf,CMS:2022dwd}.

In contrast,
there is scarce information on the exact CP nature of the $h_{125}$ couplings.
The CP nature of $h_{125} VV$ couplings, where $VV=W^+W^-, ZZ$ are a pair of weak gauge bosons,
has been probed by ATLAS in \cite{ATLAS:2023mqy}.
The CP-odd character of $h_{125} \tau\tau$ have been studied in \cite{CMS:2021sdq,ATLAS:2022akr},
while CP violation in $h_{125}tt$ has been probed in \cite{Frederix:2011zi,Broggio:2017oyu}.
All these limits still allow for substantial CP-odd components in the respective
couplings.
Notice that there are currently no significant bounds on $h_{125}bb$ couplings.
These are very active and important lines of phenomenological research.

It turns out that a singularly curious possibility was identified
in a complex two Higgs doublet model (C2HDM) with explicit CP violation:
the possibility that $h_{125}$ couples to top quarks as a pure scalar,
while it couples to bottom quarks as a pure pseudoscalar \cite{Fontes:2015mea}.
As a proof of LHC's continued importance, this possibility became disfavored
also through a combination of LHC with 
new results on the electron electric dipole moment \cite{ACME:2018yjb,Roussy:2022cmp}.
Since this tantalizing possibility was disfavored in the C2HDM, Ref.~\cite{Boto:2024jgj} returned to the issue for the complex three Higgs doublet model (C3HDM), in the simplified context of a $\Z2\times\Z2'$ symmetry. Indeed,
substantial portions of a CP-odd $bb$ couplings
can be recovered by considering a complex three Higgs doublet model
(C3HDM) with explicit CP violation \cite{Boto:2024jgj}.
Regardless,
probing the CP-odd components of $h_{125}$ is mandatory;
models of physics beyond the SM (BSM) may yield correlations with other observables and
point the searches in new direction.

Unfortunately, 
exploring the parameter space becomes increasingly challenging due to
stringent experimental constraints and the difficulty of identifying
viable regions in high-dimensional spaces. Traditional methods exhibit
low efficiency and often require assumptions that may leave significant
regions unexplored.
In the specific case of the C3HDM with traditional methods,
a random scan is not able to find experimentally viable points
in the 20 parameter space.
They are only found after making an assumption that points can
be found near the alignment limit in the real case  \cite{Boto:2024jgj}.
Recently, Machine Learning (ML) has become an essential tool
in particle physics,
driven by the increasing complexity and volume of data from experiments
like those at the LHC.
Traditional analysis techniques are being augmented or even
replaced by powerful ML algorithms that excel in pattern recognition,
anomaly detection, high-dimensional classification,
accelerated simulation and beyond the Standard Model (BSM)
parameter estimation
\cite{CrispimRomao:2020ucc,deSouza:2022uhk,Romao:2024gjx,Diaz:2024yfu,AbdusSalam:2024obf,Batra:2025amk}.
We looked into an Evolutionary Strategy Algorithm, which differs
from a classification algorithm in the sense that it finds valid
and verifiable points of the model, first introduced in \cite{deSouza:2022uhk},
combined with an anomaly detection used for Novelty
Reward developed in \cite{Romao:2024gjx} to ensure good exploration of parameter spaces, and
(most importantly) of physical consequences of a model.

This work is devoted to the use of ML techniques for a full
exploration of phenomenological consequences of new physics models.
We have found that one is able to produce model parameters in concordance with all known experimental constraints, at orders of magnitude faster rates.
And, most importantly, one is also able to find dramatic new phenomenological
features, helping both in understanding the theoretical characteristics of a model,
and also in guiding experimental searches.
Although we apply this here to a study of CP-odd Higgs couplings in a particular
context, the techniques devised here find applicability in all models;
being especially impactful in models with large numbers of parameters.

Our work is organized as follows.
In section~\ref{sec:pot}, we present the definitions adopted
for the $\Z2\times\Z2'$ symmetric C3HDM.
Section~\ref{sec:constraints} details the theoretical and experimental
constraints applied in our simulations.
In section~\ref{sec:ml} we describe the machine learning techniques
employed to explore the model’s parameter space.
Section~\ref{sec:results} presents and analyzes the results,
followed by our conclusions in section~\ref{sec:conclusions}.

\section{The C3HDM}\label{sec:pot}

\subsection{The scalar potential}

The scalar potential obeying the $\Z2\times \Z2'$ symmetry,
including soft breaking terms, is given by \cite{Botella:1994cs}
\begin{equation} \label{VNHDM}
	V = V_2 + V_4 = \mu_{ij} (\Phi_i^\dag \Phi_j) + z_{ijkl} (\Phi_i^\dag \Phi_j)(\Phi_k^\dag \Phi_l) \;,
\end{equation}
with
\begin{equation}
	V_2 =\, \mu_{11} (\Phi_1^\dag \Phi_1) + \mu_{22} (\Phi_2^\dag \Phi_2) + \mu_{33} (\Phi_3^\dag \Phi_3) + \left( \mu_{12}  (\Phi_1^\dag \Phi_2) + \mu_{13}  (\Phi_1^\dag \Phi_3) + \mu_{23}  (\Phi_2^\dag \Phi_3) + h.c.\right)
	\;, \\
\end{equation}
where $\mu_{11}$, $\mu_{22}$, $\mu_{33}$ are real, and
the complex $\mu_{12}$, $\mu_{13}$, $\mu_{23}$ parameters break the
$\Z2\times \Z2'$ symmetry softly. Moreover \cite{Boto:2022uwv},
\begin{equation}
\begin{split}
		V_4 =\,& V_{RI} + V_{\Z2\times \Z2'} \;,\\
		V_{RI} =\,& \lambda_1 (\Phi_1^\dag \Phi_1)^2 + \lambda_2 (\Phi_2^\dag \Phi_2)^2 + \lambda_3 (\Phi_3^\dag \Phi_3)^2  + \lambda_4 (\Phi_1^\dag \Phi_1)(\Phi_2^\dag \Phi_2) + \lambda_5 (\Phi_1^\dag \Phi_1)(\Phi_3^\dag \Phi_3) \\&+ \lambda_6 (\Phi_2^\dag \Phi_2)(\Phi_3^\dag \Phi_3)  + \lambda_7 (\Phi_1^\dag \Phi_2)(\Phi_2^\dag \Phi_1) + \lambda_8 (\Phi_1^\dag \Phi_3)(\Phi_3^\dag \Phi_1) + \lambda_9 (\Phi_2^\dag \Phi_3)(\Phi_3^\dag \Phi_2) \;, \\
		V_{\Z2\times \Z2'} =\,& \lambda_{10} (\Phi_1^\dag \Phi_2)^2 + \lambda_{11} (\Phi_1^\dag \Phi_3)^2 + \lambda_{12} (\Phi_2^\dag \Phi_3)^2 + h.c.
		\;,
\end{split} 
\end{equation} 
where $\lambda_1,\, ..., \lambda_9$ are real parameters and $\lambda_{10}$, $\lambda_{11}$, $\lambda_{12}$ are potentially complex. The existence of complex quadratic and quartic parameters enables explicit CP violation.  Without loss of generality, the scalar field vacuum expectation values (vevs) can be chosen as
 \begin{equation}
	\Phi_i = 
	\begin{pmatrix}
		w_i^+ \\ (v_i + x_i + i\ z_i) /\sqrt{2}
	\end{pmatrix} 
	\;, i=1,2,3,
\end{equation}
where $v_1, v_2, v_3$ are real, non-negative and related by
\beq \label{v246}
v\equiv (v_1^2+v_2^2+v_3^2)^{1/2}=\frac{2 m_W}{g}=(\sqrt{2}G_F)^{-1/2}\simeq 246.219~{\rm GeV}\,.
\eeq
The five equations obtained by minimizing the scalar potential with respect to the unit vectors of the vevs are, 
\small
\begin{equation}\begin{split}\label{STAT}
		\mu_{11} v_1 = -\Re(\mu_{12}) v_2 -\Re(\mu_{13}) v_3 - &v_1 \left( \lambda_1 v_1^2 + \left(\Re(\lambda_{10}) + \frac{1}{2}\lambda_4 + \frac{1}{2}\lambda_7\right) v_2^2 + \left(\Re(\lambda_{11}) + \frac{1}{2} \lambda_5 + \frac{1}{2}\lambda_8\right) v_3^2 \right ) 
		\;, \\*[1mm]
		\mu_{22} v_2 =  -\Re(\mu_{12}) v_1 -\Re(\mu_{23}) v_3 - &v_2 \left( \lambda_2 v_2^2 + \left(\Re(\lambda_{10}) + \frac{1}{2}\lambda_4 + \frac{1}{2}\lambda_7\right) v_1^2 + \left(\Re(\lambda_{12}) + \frac{1}{2}\lambda_6 + \frac{1}{2}\lambda_9\right) v_3^2 \right) 
		\;, \\*[1mm]
		\mu_{33} v_3 =  -\Re(\mu_{13}) v_1 -\Re(\mu_{23}) v_2 - &v_3 \left( \lambda_3 v_3^2 + \left(\Re(\lambda_{11}) + \frac{1}{2} \lambda_5 + \frac{1}{2}\lambda_8\right) v_1^2 + \left(\Re(\lambda_{12}) + \frac{1}{2}\lambda_6 + \frac{1}{2}\lambda_9\right) v_2^2\right) 
		\;, \\*[1mm]
		\Im(\mu_{13}) v_3 &=  -v_1\left( \Im(\lambda_{10}) v_2^2 + \Im(\lambda_{11}) v_3^2 \right) - \Im(\mu_{12}) v_2 
		\;, \\*[1mm]
		\Im(\mu_{23}) v_3 &= v_2\left( \Im(\lambda_{10}) v_1^2 - \Im(\lambda_{12})v_3^2 \right) + \Im(\mu_{12}) v_1
		\;.
	\end{split}
\end{equation}
\normalsize
 With the scalar minimum established, the Higgs basis \cite{Georgi:1978ri,Donoghue:1978cj,Botella:1994cs} can be introduced  by the following rotation,
\begin{equation}\label{higgsbasisZ3}
     \begin{pmatrix} H_0 \\ R_1 \\ R_2 \end{pmatrix}
     =
    R_H
     \begin{pmatrix} x_1 \\ x_2 \\ x_3 \end{pmatrix}
     =
     \begin{pmatrix} c_{\beta_2} c_{\beta_1} & c_{\beta_2} s_{\beta_1} & s_{\beta_2} \\
                    -s_{\beta_1} & c_{\beta_1} & 0 \\ 
                    -c_{\beta_1} s_{\beta_2} & -s_{\beta_1} s_{\beta_2} & c_{\beta_2}\end{pmatrix}
     \begin{pmatrix} x_1 \\ x_2 \\ x_3 \end{pmatrix},
\end{equation}
which corresponds to parameterizing the vevs as,
\begin{equation}\label{3hdmvevs}
     v_1=v c _{\beta_1} c _{\beta_2}\,,\qquad v_2=v s _{\beta_1} c _{\beta_2}\, ,\qquad v_3=v s _{\beta_2},
\end{equation}
where $c_{\beta_i}$ and $s_{\beta_i}$ account for $\cos(\beta_i)$ and $\sin(\beta_i)$, respectively.
We have 24 parameters in the scalar potential,
besides the three real vevs. These are constrained by the five stationarity conditions, yielding 22 parameters. Of these, two will be reserved for $v$ and the mass of the $125\,\textrm{GeV}$
scalar, leaving 20 free parameters.

\subsection{Physical basis\label{sec:physbasis}}
For obtaining the physical quantities we follow the procedure developed in \cite{Boto:2024jgj}. 
We start by changing $w^+_i$ and $z_i$ into the Higgs basis, by
\begin{equation}\label{e:field_transf}
	\begin{pmatrix}
		x \\ z^\prime
	\end{pmatrix}
	= \begin{pmatrix}
		\one & 0 \\
		0 & R_H
	\end{pmatrix}
	\begin{pmatrix}
		x \\ z
	\end{pmatrix}
	\;,
	\qquad \qquad
	w^{+\prime} = R_H w^+
	\; .
\end{equation}
This isolates the would be Goldstone bosons, $G^0$ and $G^\pm$. Thus, only a $2\times 2$ charged Higgs mass matrix and a $5\times 5$ neutral Higgs mass matrix remain to be diagonalized. This is performed with orthogonal $W$ and $R$ matrices (respectively) according to 
\begin{equation}\label{mass_eigenstates}
	\Big(H_1^+ \; H_2^+\Big)^T = W\;\Big(w_2^{+\prime}\; w_3^{+\prime}\Big)^T  \;, \qquad \quad \Big(h_1\;h_2\;h_3\;h_4\;h_5\Big)^T = R 
	\;\Big(x_1\;x_2\;x_3\;z_2^\prime\; z_3^\prime\Big)^T \;.
\end{equation}
We parametrize $W$ as
\begin{equation}
	W = 
	\begin{pmatrix}
		c_\theta e^{i \varphi} & s_\theta e^{-i \varphi} \\*[1mm]
		-s_\theta e^{i \varphi} & c_\theta e^{-i \varphi}
	\end{pmatrix}.
\end{equation}
In order to parametrize $R$, we start by defining the $O_{ij}$ $5 \times 5$ matrices which are like the identity matrix in all entries
except the entries $ii$ and $jj$, given by $\cos{\alpha_{ij}}$,
the entry $ij$, given by $\sin{\alpha_{ij}}$,
and the entry $ji$ given by $-\sin{\alpha_{ij}}$.
For our scans we choose
\begin{eqnarray}
R_x &=& O_{23} O_{13}O_{12} \;,
\nonumber\\
R_z &=& O_{45} \;,
\nonumber\\
R_{CPV} &=& O_{35}O_{34}O_{25}O_{24}O_{15}O_{14}\, ,\nonumber\\
R &=& R_{CPV} R_x R_z = R_{CPV} R_z R_x \; .
\label{eq:Rspecific}
\end{eqnarray}
With these definitions, taking the real limit corresponds to setting the $R_{CPV}$ as the identity matrix and setting $\varphi=0$.

For later use, we define the full set of neutral mass eigenstates as \cite{Boto:2024jgj}
\begin{equation}\label{matrixQ}
    \begin{split}
    \begin{pmatrix}
			\xi_1 \\  \xi_2 \\ \xi_3 \\ \xi_4 \\ \xi_5 \\ \xi_6
		\end{pmatrix}
        \equiv
		\begin{pmatrix}
			G^0 \\ h_1 \\ h_2 \\ h_3 \\ h_4 \\ h_5
		\end{pmatrix}
        		&=
		Q
		\begin{pmatrix}
			x_1 \\ x_2 \\ x_3 \\ z_1 \\ z_2 \\ z_3
		\end{pmatrix}
\\
	&= 		\left(\begin{matrix}0 & 0 & 0 & c_{\beta_1} c_{\beta_2} & c_{\beta_2} s_{\beta_1} & s_{\beta_2}\\R_{11} & R_{12} & R_{13} & - R_{14} s_{\beta_1} - R_{15} c_{\beta_1} s_{\beta_2} & R_{14} c_{\beta_1} - R_{15} s_{\beta_1} s_{\beta_2} & R_{15} c_{\beta_2}\\R_{21} & R_{22} & R_{23} & - R_{24} s_{\beta_1} - R_{25} c_{\beta_1} s_{\beta_2} & R_{24} c_{\beta_1} - R_{25} s_{\beta_1} s_{\beta_2} & R_{25} c_{\beta_2}\\R_{31} & R_{32} & R_{33} & - R_{34} s_{\beta_1} - R_{35} c_{\beta_1} s_{\beta_2} & R_{34} c_{\beta_1} - R_{35} s_{\beta_1} s_{\beta_2} & R_{35} c_{\beta_2}\\R_{41} & R_{42} & R_{43} & - R_{44} s_{\beta_1} - R_{45} c_{\beta_1} s_{\beta_2} & R_{44} c_{\beta_1} - R_{45} s_{\beta_1} s_{\beta_2} & R_{45} c_{\beta_2}\\R_{51} & R_{52} & R_{53} & - R_{54} s_{\beta_1} - R_{55} c_{\beta_1} s_{\beta_2} & R_{54} c_{\beta_1} - R_{55} s_{\beta_1} s_{\beta_2} & R_{55} c_{\beta_2}\end{matrix}\right)
		\begin{pmatrix}
			x_1 \\ x_2 \\ x_3 \\ z_1 \\ z_2 \\ z_3
		\end{pmatrix} .
	\end{split}
\end{equation}

Taking $\xi_2\equiv h_1$ as the  $125\,\textrm{GeV}$ Higgs, it is easy to identify the
conditions that guarantee that $h_1$ couples as the SM Higgs.
These conditions may be written as
\begin{equation}
R_{1k}=\left(R_H\right)_{1k}\,,\quad (k=1,2,3)\,,\qquad \, R_{14}=R_{15}=0\,,
\end{equation}
or, which is the same, as
\begin{equation}
\alpha_{12}=\beta_1\,,\quad \alpha_{13}=\beta_2\, ,\quad \alpha_{14}=\alpha_{15}=0\,.
\end{equation}
These conditions align the $125\,\textrm{GeV}$ Higgs regardless of the exact values of $ \alpha_{23},\, \alpha_{24},\,\alpha_{25},\,
\alpha_{34},\, \alpha_{35},\,\alpha_{45}$.
Indeed, one can show that the $hVV$ coupling divided
by its SM value ($k_V$) may be written as
\begin{equation}
    \frac{k_V}{\cos{\alpha_{14}}\cos{\alpha_{15}}} = 1-2\sin{\left(\frac{\alpha_{12}-\beta_1}{2}\right)^2}\cos{\left(\frac{\alpha_{13}+\beta_2}{2}\right)^2}-2\sin{\left(\frac{\alpha_{13}-\beta_2}{2}\right)^2}\cos{\left(\frac{\alpha_{12}-\beta_1}{2}\right)^2}\,.\label{eq:k_V}
\end{equation}
\subsection{Independent parameters\label{sec:parameters}}

Ideally one would like to perform
an extensive scan of the parameter space.
Our fixed inputs are $v = 246\,\textrm{GeV}$ and $m_{h_1} = 125\,\textrm{GeV}$. We then would take random values in the
ranges:

\begin{align}
  &\theta,\,\varphi,\, \in
\left[-\pi,\pi\right]; \label{eq:scanparameters1}\\[8pt]
& \alpha_{12},\, \alpha_{13},\, \alpha_{14},\,
\alpha_{15},\, \alpha_{23},\, \alpha_{24},\,\alpha_{25},\,
\alpha_{34},\, \alpha_{35},\,\alpha_{45}\, \in
\left[-\pi,\pi\right];\\[8pt]
&\tan{\beta_1},\,\tan{\beta_2}\,\in \left[0.3,10\right]; 
\\[8pt]
& m_{h_2}\, 
\in \left[126,1000\right]\,\textrm{GeV},\,
m_{H_1^\pm},\,m_{H_2^\pm}\,
\in \left[100,1000\right]\,\textrm{GeV};\\[8pt]
&
\text{Re}(m^2_{12}),\text{Re}(m^2_{13}),
\text{Re}(m^2_{23}) \in  \left[\pm 10^{-1},\pm 10^{7}\right]\,
\textrm{GeV}^2\, .
\label{eq:scanparameters5}
\end{align}
These 20 free parameters (in addition to $v$ and $m_{h_1}$), fully determine the point in parameter space, as we had found from the Lagrangian. The masses $m_{h_3},\, m_{h_4},\, m_{h_5}$ are noticeably absent from the previous list. This occurs because they are not independent from the parameters in 
Eqs.~\eqref{eq:scanparameters1}-\eqref{eq:scanparameters5}; indeed this model has the constraint equations
\begin{equation}\label{e:X1i}
	m_{h_i}^2 \left[ R_{i5} c_{\beta_2}(R_{i1}s_{\beta_1}-R_{i2}c_{\beta_1}) - R_{i3}R_{i4} \right] = X_{1i} m_{h_i}^2 = 0
	\;,
\end{equation}
\begin{equation}\label{e:X2i}
	m_{h_i}^2 R_{i5} \frac{ c_{\beta_2}(R_{i1}c_{\beta_1}+R_{i2}s_{\beta_1}) - R_{i3} s_{\beta_2} }{s_{\beta_2}}  = X_{2i} m_{h_i}^2  = 0
	\;,
\end{equation}
\begin{equation}\label{e:X3i}
	m_{h_i}^2 \frac{ R_{i4} (R_{i1}c_{\beta_1}-R_{i2}s_{\beta_1}) - R_{i5} s_{\beta_2}(R_{i1}s_{\beta_1}+R_{i2}c_{\beta_1}) }{s_{\beta_1}}  = X_{3i} m_{h_i}^2  = 0
	\;,
\end{equation}
where we have implicitly defined a $3\times 5$ matrix $X$. Inverting the system and assuming $m_{h_1}^2$ and $m_{h_2}^2$ are given, 
\begin{equation}
	\begin{split}
		m_{h_3}^2 = 
		-\frac{ \sum_{i=1}^2 (X_{1i}X_{24}X_{35}-X_{1i}X_{25}X_{34}-X_{14}X_{2i}X_{35}+X_{14}X_{25}X_{3i}+X_{15}X_{2i}X_{34}-X_{15}X_{24}X_{3i}) m_{h_i}^2}{X_{13}X_{24}X_{35}-X_{13}X_{25}X_{34}-X_{14}X_{23}X_{35}+X_{14}X_{25}X_{33}+X_{15}X_{23}X_{34}-X_{15}X_{24}X_{33}} 
		\;,
	\end{split}\label{mh3}
\end{equation}
\begin{equation}
	m_{h_4}^2 = - \frac{1}{X_{24}X_{35}-X_{25}X_{34}} \sum_{i=1}^3 (X_{2i}X_{35}- X_{3i} X_{25}) m_{h_i}^2
	\;,\label{mh4}
\end{equation}
\begin{equation}
	m_{h_5}^2 = -\frac{1}{X_{35}}\sum_{i=1}^4  X_{3i}  m_{h_i}^2
	\;.\label{mh5}
\end{equation}

Notice that Eqs.~\eqref{mh3}-\eqref{mh5} do \textit{not} guarantee that the
left-hand sides are indeed positive.
Thus, points in parameter space yielding negative squared masses are to
be discarded \footnote{The non-generality of the mass matrices is similar to what happens
in the C2HDM, where there is a single constraint.
However, the presence of one single constraint on the C2HDM makes it
easily solvable for one of the rotation angles, unlike in our C3HDM,
where the situation is much more complicated.}.

Explicit expressions for the quartic potential parameters $\lambda_{1-12}$ and the remaining quadratic terms, when written in terms of the parameters in Eqs.~\eqref{eq:scanparameters1}-\eqref{eq:scanparameters5}, can be found in \cite{Boto:2024jgj}.

\subsection{The Yukawa Lagrangian} \label{ssub:yukawa}

 The imposed $\Z2\times \Z2'$ symmetry acts on the scalars fields, right-handed down quarks ($d_R$) and right-handed charged-leptons ($\ell_R$) as 
\begin{equation}
    \begin{aligned}
     \label{eq:Symmetry}
     \Z2 :&\ \Phi_1 \to -\Phi_1,\quad \ell_R \rightarrow -\ell_R, \\
      \Z2' :&\ \Phi_2 \to - \Phi_2,\quad d_R \rightarrow -d_R ,
    \end{aligned}
\end{equation}
resulting in the type-Z couplings of the model, in which each scalar couples to a different family of fermions. In this configuration, one may assume the right-handed fermions of each electric-charge to couple only to one of the Higgs doublets, which will be dubbed $\Phi_u$, $\Phi_d$ and $\Phi_\ell$ going forward. To understand how the neutral scalars couple to fermions, we need to look at the terms of the Yukawa Lagrangian, given by, 
%
\begin{equation}\label{LY}
- \mathcal{L}_\text{Yukawa} = \overline{Q}_L Y^u \Phi_d n_R
+\overline{Q}_L  Y^d \tilde{\Phi}_u p_R
+\overline{L}_L Y^\ell \Phi_\ell \ell_R + h.c. 
\;,
\end{equation}
where $Q_L = (p_L \; n_L)^T$, $L_L = (\nu_L \; \ell_L)^T$,
while $n_R$, $p_R$, and $\ell_R$ are,
respectively,
right-handed down-type, up-type, and charged lepton fields,
written in a weak basis. The Yukawa matrices $Y^u, Y^d$ and $Y^\ell$ are $3\times 3$ in the respective fermionic sectors.
After Spontaneous Symmetry Breaking (SSB), the scalars acquire vevs leading to mass terms for the fermions. In this work, we do not consider right-handed neutrino fields. As a result, the neutrinos will be massless, and we can choose the charged lepton basis such that the $Y^\ell$ matrix is already diagonal:
\begin{equation}
    \frac{v_\ell}{\sqrt{2}}Y^\ell = D_\ell \equiv \text{diag}(m_e, m_\mu, m_\tau).
\end{equation}
As for the quarks, to perform the required basis change from flavour to the diagonal mass basis, one may rotate the quarks fields by an unitary transformation of the form
\begin{equation}
    f_{L/R} = U^f_{L/R}\tilde{f}_{L/R}.
\end{equation}
As a result, 
%
%
\begin{equation}
      (U^{f}_L)^\dagger  \frac{v_f}{\sqrt2} Y^f U^{f}_R \equiv D_{f} ,
\end{equation}
where $f = u, d$, $D_u=\text{diag}(m_u,m_c,m_t)$, and $D_d=\text{diag}(m_d,m_s,m_b)$. By changing to the quark mass basis, we obtain the Cabibbo-Kobayashi-Maskawa (CKM) matrix \cite{bib:CKM-Cabibbo,Kobayashi:1973fv}, $V_{\text{CKM}}=(U^p_L)^\dagger U^n_L$ . 

The Lagrangian is now composed of fermion mass terms and scalar-fermion interactions, for which the CKM matrix is involved in the terms with charged scalars. The expression for the interactions with the neutral scalars, $\xi_{1-6}$, are
\begin{equation}
    -\mathcal{L}_{\xi ff} = \sum_f \sum^{2N}_{j=1} \frac{m_f}{v} \Bar{f} \left( c^e_{\xi_j ff} + i\gamma_5 c^o_{\xi_j ff} \right) f \xi_j ,
\end{equation}
where,
\begin{equation}\label{h0ff_couplings}
	\;\; c^e_{\xi_j ff} + i\gamma_5 c^o_{\xi_j ff} = \frac{v}{v_f}\left( Q_{jf} \pm i \gamma_5 Q_{j,N+f}\right)
	\;,
\end{equation}
while $f$ refers to the doublet that couples to a given fermion, the $``+''$ is applied to leptons or down-type quarks, and the $``-''$ for up-type quarks. In particular, for the $125\,\textrm{GeV}$ Higgs found at LHC,
\begin{equation}
    c^e_{h_{125}ff} + i\gamma_5c^o_{h_{125}ff} = \frac{v}{v_f} (Q_{2,f} \pm i\gamma_5 Q_{2,3+f}),
\end{equation}
or, in terms of the rotation angles,
\begin{equation}
	c^e_{ff} \equiv
	c^e_{h_{125}ff} =  
	\frac{R_{11}}{c_{\beta_2} c_{\beta_1}} \;,\;
	\frac{R_{12}}{c_{\beta_2} s_{\beta_1}} \;,\;
	\frac{R_{13}}{s_{\beta_2} } \;,\;
	\qquad \text{for} \quad f=\ell,d,u \;,
\end{equation}
and
\begin{equation}
	c^o_{ff} \equiv
	c^o_{h_{125}ff} = 
	\frac{-R_{14}s_{\beta_1}-R_{15}c_{\beta_1}s_{\beta_2}}{c_{\beta_2} c_{\beta_1}} \;, \;
	\frac{R_{14}c_{\beta_1}-R_{15}s_{\beta_1}s_{\beta_2}}{c_{\beta_2} s_{\beta_1}} \;,\;
	-  \frac{R_{15} c_{\beta_2}}{s_{\beta_2}}\, ,
	\qquad \text{for} \quad f=\ell,d,u \;. \label{eq:cff_odd}
\end{equation}

If we take the SM-like limit for the couplings,
the even components and odd components are determined to be
$c^e_{ff} = 1$ and $c^o_{ff}=0$, respectively.

There are limits on the CP-odd coupling of the $125\,\textrm{GeV}$ Higgs coupling
with the top, arising from $tth$ production
\cite{PhysRevLett.125.061802, PhysRevLett.125.061801, ATLAS:2023cbt}.
Taking  $c^o_{tt} = 0$, 
\cite{Boto:2024jgj} found
an analytical relation between the odd coupling of the $\tau$ and $b$ quarks:
\begin{equation}
    \frac{c^o_{\tau\tau}}{c^o_{bb}} = - \frac{v_2}{v_1} = - \tan^2\beta_1 \, .
\label{notthere}
\end{equation}
This would mean that,
despite the apparent uncorrelated nature of the scalar coupling to the different fermion types in this type-Z model, there would indeed exist a remnant correlation among the odd couplings in this regime.
With the simulations done at the time (which, in order to find good points,
had to start close to the real case of the alignment limit),
both $c^o_{tt} \sim 0$ and Eq.~\eqref{notthere} seemed to be
confirmed by the numerical results.
As we will see below,
one dramatic result from our new method is that we can now
produce points consistent with all known data where both these
statements are far from realized.

\section{Constraints} \label{sec:constraints}

The model consistency is equivalent to requiring a set of
theoretical and experimental constraints.
For this analysis we choose the first neutral scalar as the
$125\,\textrm{GeV}$ Higgs. As described in Sec.~\ref{sec:parameters},
the parametrization leads to the squared mass of three other neutral
scalars to be derived parameters.
The requirement that all three are positive quantities is set as a
constraint that must be satisfied before the fitting procedure
is initiated \cite{deSouza:2025uxb}.
During the simulation, all other constraints are considered on equal footing, both theoretical and experimental,  with the techniques described below.
This is a distinguishing feature of this method; (after the initial
squared mass step) there is no hierarchical sequence of constraints.
This is a very positive asset for the approach.
Indeed, for most models, one does not know ahead of the simulation which particular observables will be easier to obey, and which will turn out to be very difficult and require a dedicated analysis.
With this method one can remain agnostic to such foresight.

The constraints are:

\begin{itemize}
\item
\textbf{Boundedness from Below:}
Necessary and sufficient BFB conditions are not yet known for the $\Z2\times\Z2'$ symmetric potential. We apply the sufficient conditions presented in Ref.~\cite{Boto:2024jgj}, which were obtained with the method derived in Ref.~\cite{Boto:2022uwv}, based on the copositivity conditions of \cite{Kannike_2012,Klimenko:1984qx}.
\item
\textbf{Perturbativity:} We must also ensure the perturbativity of the Yukawa couplings by setting them to be $|y_i| < \sqrt{4\pi}$, where $i = t, b, \tau$.
\item 
\textbf{Unitarity:}
The unitarity bounds for this model were obtained in section 5.10 of Ref.~\cite{Bento_2022} using the notation of Ref.~\cite{Ferreira_2008}, for which we must guarantee that the absolute value of the eigenvalues of the scattering matrices is smaller than $8\pi$ or, similarly,
that they are bounded as $|\lambda_i| < 8\pi$ for all $i=1 \cdots 27$.
\item
\textbf{Oblique Parameters STU:} In order to use the results of \cite{Grimus_2008}, we take the matrices $U$ and $V$ defined in appendix C of \cite{Boto:2024jgj} to compare with the most recent fit of \cite{Baak2014}.
\item \textbf{Flavour searches}: With FCNCs excluded by symmetry, the most significant observable corresponds to the NP contribution at one-loop order to $b \to s \gamma$ . Following the procedure described in Refs.~\cite{Florentino:2021ybj,Boto:2021qgu, Akeroyd:2020nfj}, we impose the $3\sigma$ experimental limit,
\begin{equation}
\label{e:b2sg}
2.87 \times 10^{-4} < \text{BR}(B\to X_s \gamma) < 3.77 \times 10^{-4}\,.
\end{equation}
Ref.~\cite{Chakraborti:2021bpy} found that the constraints coming from the meson mass differences correspond to excluding very low values of $\tan\beta_{1,2}$. Thus, when setting our parameter domain we imposed
\begin{eqnarray}
	\tan\beta_{1,2} > 0.3 \, .
\end{eqnarray}
\item 
\textbf{eEDM:}  The calculation of the non-zero electric dipole moment of the electron
(eEDM) follows the formulae in
\cite{Barr:1990vd,Yamanaka:2013pfn,Abe:2013qla,Inoue:2014nva,Altmannshofer:2020shb}, with significant experimental constraints  \cite{ACME:2018yjb,Roussy:2022cmp}. The contributions from the muon EDM and from non-leptonic EDMs are currently less stringent~\cite{Fontes:2017zfn} and will not be considered here. 
\item \textbf{LHC signal strengths:}
We comply with the most recent collider measurements on the $125 \textrm{GeV}$ Higgs.
By combining all the relevant
production and decay channels, $pp \to h \to f$,
the Higgs signal strengths $\mu_i^f$ are defined as,
\begin{equation}
	\mu_i^f = \frac{\sigma^{\text{3HDM}}_{i}(pp \rightarrow h_{125})}{\sigma^{\text{SM}}_{i}(pp \rightarrow h_{125})} \times\frac{\text{BR}^{\text{3HDM}}(h_{125} \rightarrow f)}{\text{BR}^{\text{SM}}(h_{125} \rightarrow f)}\,,
\end{equation}
with the subscript `$i$' indicating the production mode,
and the superscript `$f$' for the decay channel of the $125\,\textrm{GeV}$ Higgs.
For the relevant production mechanisms of the neutral scalars,
we calculate the gluon fusion ($ggF$) cross section with HIGLU~\cite{Spira:1995mt}
at NNLO, with the effect of the bottom quark included. For the vector boson fusion (VBF)
and associated production with a vector boson ($VH$, $V = W$ or $Z$), $k_V^2$
gives the ratios of the model cross section with respect to the SM,
obtained from \url{https://twiki.cern.ch/twiki/bin/view/LHCPhysics/LHCHWGCrossSectionsFigures}.
For the associated production with a pair of top quarks ($ttH$),
the pure CP even and the pure CP odd cross sections are
different~\cite{Frederix:2011zi,Broggio:2017oyu},
so the cross section for the model is obtained by multiplying the
SM result by,
\begin{equation}
	\label{eq:7}
	(c_{tt}^e)^2+  \frac{\sigma^{\rm ttH}_{0^-}}{\sigma^{\rm ttH}_{0^+}}\
	(c_{tt}^o)^2  \, ;
\end{equation}
Ref.\cite{Broggio:2017oyu} provides a value
$\sigma^{\rm ttH}_{0^-}/\sigma^{\rm ttH}_{0^+} \simeq 0.416$,
at NLO+NLL order, after setting  $M_H=125$ GeV and $\sqrt{s}=13$ TeV.
Our black-box code combines the production cross sections with the calculation
of all the decays using the Feynman rules derived with Feynmaster
\cite{Fontes:2021iue,Fontes:2025svw}. The loop decay $h\to\gamma\gamma$
is calculated by adapting our previous results \cite{Fontes:2014xva}.
We demand that all signal strengths have a $2\sigma$ agreement with the
most recent ATLAS results \cite{ATLAS:2022vkf},
which are also consistent with CMS \cite{CMS:2022dwd}.
As a cross-check to our in-house program, we also used the \texttt{HiggsSignals} module
of \texttt{HiggsTools-1.1.3} at $\Delta \chi^2$ corresponding to $3\sigma$.
\item  
\textbf{Searches for new particles:}
Direct searches at the LHC for the new particles are considered
using the \texttt{HiggsBounds} package in \texttt{HiggsTools-1.1.3},
which provides a complete list of relevant experimental searches.
The production of charged scalars
in association with $t$ quark, $pp\to H^\pm t\,(b)$ , is calculated
with the routine in \texttt{HiggsTools-1.1.3}, for the $13\,\textrm{TeV}$
LHC and mass range $145\,\textrm{GeV}-2\,\textrm{TeV}$. For values below 145 GeV, HiggsTools-1.1.3 estimates the cross section from the $BR(t \to H^+ b)$ that we give to the program.
\item
\textbf{Direct searches for CP-violation:}
There are CP-violation constraints
from the decays of the $125 \textrm{GeV}$ Higgs into $\tau \bar{\tau}$
\cite{CMS:2021sdq,ATLAS:2022akr}:
$|\theta_{\tau}| = |\arctan(c^o_{\tau} / c^e_{\tau})| < 34^\circ$.
There are also limits on the CP-odd coupling of the $125\,\textrm{GeV}$ Higgs coupling
with the top, arising from $tth$ production
\cite{PhysRevLett.125.061802, PhysRevLett.125.061801, ATLAS:2023cbt}.
We show the $1\sigma$ and $2\sigma$ contour plots in the relevant figures.
\end{itemize}

\section{Black-box Machine Learning Optimisation} \label{sec:ml}

To explore the 20 parameters with the full domain given in Eq.~\eqref{eq:scanparameters5}, we adapted the AI black box optimisation approach first presented in \cite{deSouza:2022uhk} and applied to a real 3HDM in \cite{Romao:2024gjx}.
The procedure starts with defining the constraint function, $C(\mathcal{O})$, as
\begin{equation}
	C(\mathcal{O}) = \max(0, -\mathcal{O} + \mathcal{O}_{LB}, \mathcal{O} - \mathcal{O}_{UB}),
\end{equation}
where $\mathcal{O}$ is the value of an observable or a constrained quantity,
$\mathcal{O}_{LB}$ is its lower bound, and $\mathcal{O}_{UB}$ its upper bound.
The constraint function $C(\mathcal{O})$ returns 0 only if $\mathcal{O}$
lies within the respective interval; otherwise, it returns a positive
number quantifying \textit{how far} the value of the observable is
from the defined bounds.
Importantly, points that fail to satisfy constraints are not wasted, since they still yield useful information.
The values $\mathcal{O}$ are obtained by a computational routine that
takes a vector $\theta$ in the parameter space as inputs to calculate physical quantities $\mathcal{O}(\theta)$,
including the observables described in Section~\ref{sec:constraints}.
This routine is treated as a black box, with the calculation unaffected
by the optimisation procedure.
Finding points that satisfy a given constraint is equivalent to minimizing the respective  $C(\mathcal{O})$ function. To address multiple constraints simultaneously, we make the choice of optmising a \textit{single} loss function that encapsulates all the constraints of the model as
\begin{equation} \label{eq:lossf}
	L(\theta) = \sum_{i=1}^{N_c} C(\mathcal{O}_i(\theta)),
\end{equation}
where the sum runs over all the $N_c$ constraints,
ensuring $L \geq 0$ for all vectors $\theta$ in the parameter space,
with $L = 0$ only when all constraints are
satisfied \footnote{In terms of implementation details, we perform two further 
transformations before using Eq.~\eqref{eq:lossf}  as the target function. First, we compute
the logarithm of the different $\mathcal{C}$ contributions to set them all to equivalent
nominal values and to endow us with a notion of numerical infinite necessary to
weed out unphysical points. Secondly, each $C$  contribution is min-max normalised
within each generation so that no constraint wins over any other.}.
Notably, the quantity $\mathcal{O}_i$ does not necessarily need to be an experimental observable.
Allowing the mixture of measurements, theoretical constraints and cuts in the same loss
function is a key strength of this methodology. While $\mathcal{O}_i$
may be referred to as ``observables'' in subsequent discussions,
we emphasize the flexibility to select diverse constraint types,
which allowed the exploration of exotic couplings.
Given specified bounds, the method efficiently
attempts to find valid points in parameter space, without knowledge
of the form of the likelihood. A $\Delta \chi^2$ test is performed
by calculating a single quantity and imposing a numerical bound
corresponding to $3\sigma$.
We used the \texttt{HiggsSignals} module of \texttt{HiggsTools-1.1.3} for
this calculation involving the $125 \textrm{GeV}$ Higgs couplings.

\subsection{Optimisation with an Evolutionary Strategy}

Evolutionary Strategies (ES) are a subset of optimisation algorithms of the
broader field of Evolutionary Computing, which in turn cover algorithms
based on the principals of natural selection.
Evolutionary Computing algorithms all share the
following characteristics/components: a population (set of candidate
solutions to the optimisation problem, which in our case are points
in the parameter space); a recombination procedure where a new population
is generated from previous ones, a mutation step where a population is
\textit{perturbed} to increase diversity; a fitness measure (which in our case
is the loss function) that is used to prioritize what characteristics are
inherited by new populations; and, an iteration loop over generations of populations.
In ES the optimisation takes place with an iterative process that samples
the best points as candidate solutions for a problem, extracting them from
a distribution and using them to generate the new distribution,
from which the next generation is sampled.

The sampling algorithm is the
Covariant Matrix Adaptation Evolutionary Strategy (CMA-ES)
\cite{Hansen2001,hansen2023cmaevolutionstrategytutorial},
implemented in a python package \cite{nomura2024cmaessimplepractical}.
In CMA-ES, the distribution is a highly localized multivariate normal,
initialized with its mean at a random point in parameter space and its
covariance matrix set to the identity matrix, $\one$, scaled by a constant,
$\sigma$. A generation of candidate solutions is sampled from this distribution,
and their losses are evaluated using Eq.~\eqref{eq:lossf}.
The candidates are ranked from best to worst, meaning how close
they are to fulfilling the constraints, with the best candidates used to
compute a new mean and approximate the covariance matrix.

This change in mean directs the algorithm towards the steepest descent of the loss function, similar to a gradient descent as an example of a first-order method, while the covariance matrix approximates the local Hessian, similar to a second-order method. However, CMA-ES does not compute derivatives, making it suitable for not so well behaved loss functions, which allows it to converge rapidly across a variety of optimisation problems.

\subsection{Novelty Reward} \label{ssub:NoveltyReward}

Although CMA-ES converges quickly, its reliance on a single multivariate normal exhibits limited discovery capacity due to the localized nature of its solution distribution. To address this, \cite{Romao:2024gjx} introduced a \textit{novelty reward} into the loss function, penalizing solutions based on the density of previously identified valid regions. This density is obtained using a Histogram-Based Outlier Score (HBOS), a univariate anomaly detection algorithm \cite{HBOS}. HBOS constructs a histogram for each parameter dimension in the parameter space $\mathcal{P}$  and/or observable space $\mathcal{O}$, dividing the numerical data into equal-width bins over the range of allowed values. The frequency of samples that fall into each bin is used as an estimate for the density, given by the height of the bins. The penalties are normalised to $p \in [0,1]$, such that novelty points receive a penalty of value 0, corresponding to a lower density, and points similar to previously explored approach a maximal penalty of 1.

This adjustment encourages CMA-ES to explore novel regions by penalising the loss function when constraints are satisfied but solutions are near already discovered areas.

To ensure proper minimization with these penalties, the loss function is shifted accordingly,
\begin{equation}
	\Tilde{L}(\theta) =
	\left\{
	\begin{array}{ll}
		1 + L(\theta)   & \text{if } L(\theta) > 0 \\
		\text{ } \\
		0               & \text{if } L(\theta) = 0
	\end{array}
	\right..
\label{eq:4545}
\end{equation}
We can then add the penalties and obtain the final version of the loss function,
\begin{equation}
	L_T(\theta) = \Tilde{L}(\theta) + \frac{1}{2} \left( \frac{1}{N^{\mathcal{P}}_p} \sum_{i=1}^{N^{\mathcal{P}}_p} p^{\mathcal{P}}_i(\theta^i) + \frac{1}{N^{\mathcal{O}}_p}\sum_{i=1}^{N^{\mathcal{O}}_p}p^{\mathcal{O}}_i [\mathcal{O}^{i}(\theta)]\right),
\label{eq:4646}
\end{equation}
where $p^{\mathcal{P}}_i(\theta^i)$ is the density penalty of the parameter
space $\mathcal{P}$, normalised by the number of parameter penalties
considered, $N^{\mathcal{P}}_p$, and $p^{\mathcal{O}}_i [\mathcal{O}^{i}(\theta)]$
is the density penalty of the observable space $\mathcal{O}$,
also normalised by the number of observables considered to be penalised,
given by $N^{\mathcal{O}}_p$.
For a valid point,
$\sum_i C(\mathcal{O}_i) = 0 \Rightarrow  L_T \in [0, 1]$,
and for invalid points $L_T > 1$.
We stress that penalties do not need to apply to all parameters $\theta$. One may choose a subset of parameters of interest, ${\theta^i}$, and perform \textit{focused runs} with density penalty on those specific parameters. The same applies to the space of observables $\mathcal{O}(\theta)$, with the additional benefit of achieving a direct improvement in the exploration of novel phenomenological features of a model.

By design, each run is independent, as they are initialized with new values for the CMA-ES mean and covariant matrix parameters and trained solely on points from that run. The information from previous runs can however be also used to guide new runs using seeds. We may choose valid points from previous runs, store their parameter values and start new runs initialized already in that region.
CMA-ES allows initialization with a specific mean of the multivariate Gaussian and an overall scale of the covariant matrix, $\sigma$. Unseeded runs start with $\sigma=1$, while seeded runs with  $\sigma=0.01$ are confirmed to already start at the minimum of the constraint loss function, with $\sigma=0.1$ increased in an attempt to uncover new features.

\subsection{Simulation strategy}

In previous works on 3HDMs, namely on analysis of a $\Z2\times \Z2'$ symmetric
Complex 3HDM (C3HDM), it was verified that, as in the real 3HDM case,
a simple random scan has very low probability of succeeding in finding
an allowed point \cite{Boto:2024jgj, Boto:2022uwv, Boto:2021qgu, Boto:2024tzp, Das:2019yad}.
In the C3HDM, we found less than 1 valid point per $10^{13}$ points sampled.
It was only possible to obtain the results in \cite{Boto:2024jgj} with the
adoption of a trick, in which the generated points were sampled in the real
limit of the model, closer to alignment. These should also be good C3HDM points,
which were then used as initial conditions to scan the C3HDM and enlarge
the pseudoscalar component of the points iteratively.

In this work, we adopt a very different strategy, taking advantage of the ML
methods described above. When we perform a run, CMA-ES varies the parameter
values progressively in the direction of valid points.
CMA-ES can quickly converge to valid regions in the parameter space.
However, since these regions represent global minima of the loss function
$L(\theta)$ in Eq.~\eqref{eq:lossf}, it is not able to further explore after
convergence by itself, providing little coverage of both the parameter
and the phenomenological spaces. Here, the introduction of the Novelty Detection
plays a crucial role to force the exploration in these spaces after
convergence by effectively modifying the loss function by adding
the appropriate penalty density as in Eq.~\eqref{eq:4545} and Eq.~\eqref{eq:4646}.

To force the algorithm to increase exploration for these particular parameters,
we can add them to the penalized quantities as explained in
Section~\ref{ssub:NoveltyReward}, which will then focus
the search on the chosen parameters and help populate plots.
After a successful run that uncovered promising points,
we generate seeds from them to help steer subsequent runs toward the desired regions,
improving both convergence and exploration.\\
Both features, focused runs and seeds, can and should be used
simultaneously when appropriate to boost performance.
In the results section of the paper, we present only seeded and
focused runs.
We also specify the plane variables that are being focused on,
which generally include the plotted variables.
Moreover, we launch multiple independent runs in parallel,
each using a single core, in order to enhance the overall
computational efficiency of the workflow.

We should stress that sets of points generated in the fashion proposed in this article
do \textbf{not} have a final statistical interpretation; neither a frequentist
nor a posterior interpretation.
These points were generated by \textit{focusing} on certain sets of observables
(in the technical sense explained above). What is sought in this method is a thorough exploration
of the possible phenomenological consequences of a given model, and not an attribution
of likelihood of any kind.

\section{Results and Discussion}\label{sec:results}

In this section, we present the final combination of our scans with the implemented techniques.
In several subsequent figures, we adhere to the following color code:
\begin{itemize}
\item
The points in \textbf{red} pass all the theoretical and experimental constraints described
in Section~\ref{sec:constraints}, including a $2\sigma$ agreement with
each individual $125\,\textrm{GeV}$ signal strength, and the new Higgs searches in
the \texttt{HiggsBounds} module found in \texttt{HiggsTools-1.1.3}.
In this way of generating points, there may be a large number of observables
which lie at the edge of their allowed regions, yielding a large overall $\Delta \chi^2$.
\item
The points in \textbf{green} combine points originally in red that are later found to
also satisfy $\Delta \chi^2$ at $3\sigma$, calculated using \texttt{HiggsSignals-2}
module in \texttt{HiggsTools-1.1.3}, with points generated with
$\Delta \chi^2$ at $3\sigma$ as a constraint in the machine learning algorithm,
with an appropriate  $C(\mathcal{O})$ contribution to Eq.~\eqref{eq:lossf}.
Thus, points in green imply the usual attribution of significance to the
$\Delta \chi^2$ of points on the theoretical parameter space.
\item
The points in \textbf{blue} are imported directly from \cite{Boto:2024jgj}, 
without requirements on $\Delta \chi^2$,
to draw comparisons between the methods used in \cite{Boto:2024jgj} and
the new methods proposed in this article,
serving also as a cross-check on the validity and quality
of our results.
\end{itemize}
We are able to reproduce and improve dramatically
on the work of \cite{Boto:2024jgj} for all plots shown.

The plots below consist of order $5\times 10^8$ parameter points in different projection planes,
and combine scans where the novelty reward was aimed at different sets of parameters.
The first highlight of our technique was the ability to generate valid parameter
points within the full parameter domain, and at high efficiency, with order $10^5$
points in less than 100 CPU hours.\footnote{The performance of CMA-ES
and the  Novelty Detection component has been thoroughly discussed in Ref.~\cite{Romao:2024gjx} in the
context of the 3HDM. The findings were that the overhead of the ES and
the Novelty Detection are at most of $\mathcal{O}(10\%)$ of the total computational
time which is a well worthwhile cost given the massive gain in sampling efficiency.}
This should be compared with the painstaking job performed in \cite{Boto:2024jgj}
of having to devise a strategy
of scanning around solutions of the real 3HDM and slowly
increasing the CP-violating phases.

As the signs of $c^e_{ff}$ and $c^o_{ff}$ have no meaning without taking into consideration
the sign of  $k_V \equiv c(h_{125}V V )$, in the following plots we always
present quantities in the form  $\text{sgn}(k_V)c^e_{ff}\,\text{vs. sgn}(k_V)c^o_{ff}$,
with $k_V$ given by eq.~\eqref{eq:k_V}.

\subsection{$b$ quark couplings}
\begin{figure}[htbp!]
	\centering
	\includegraphics[height=8cm]{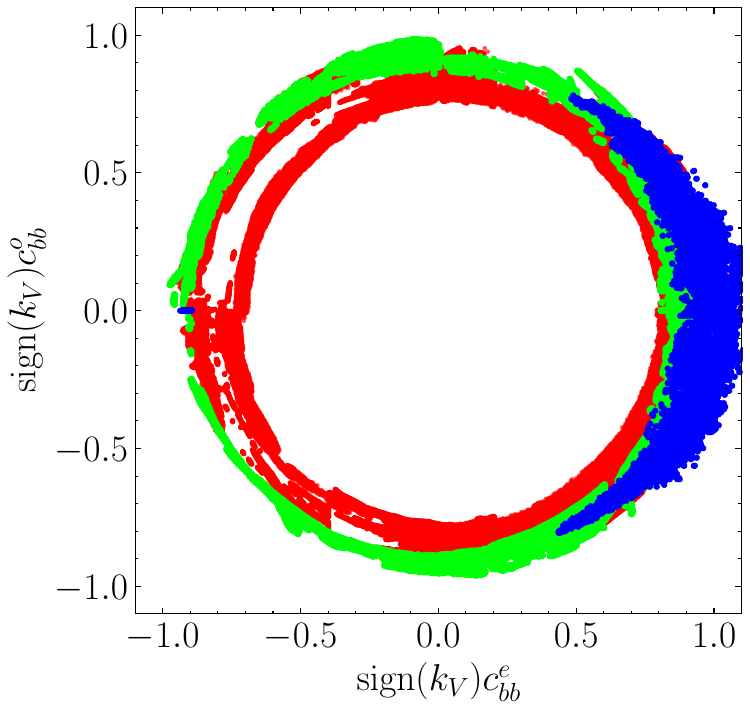}
	\hfill
	\includegraphics[height=8cm]{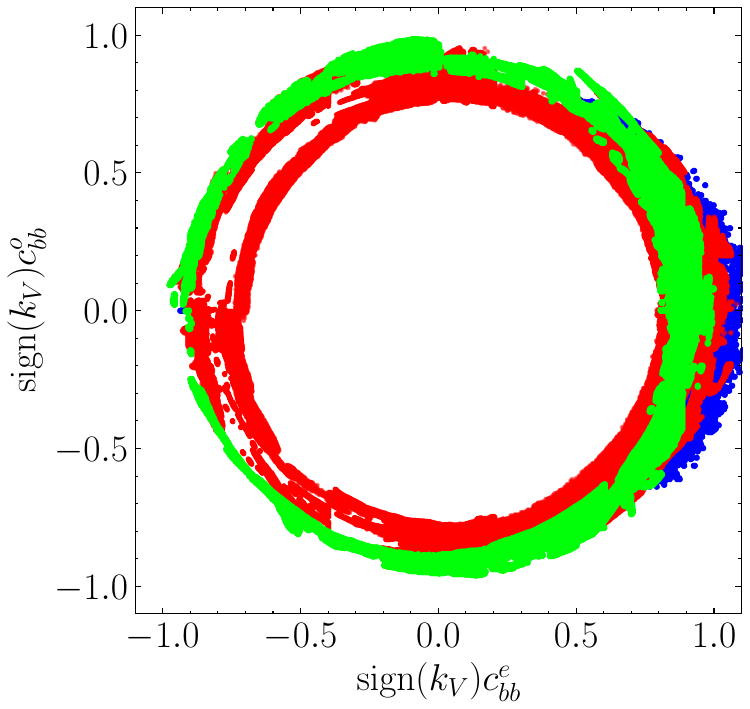}
	\caption{Combined seeded plots with CMA-ES, novelty detection and \textbf{including focus on $b\bar{b}$ couplings}. The blue coloured points from the work in \cite{Boto:2024jgj} are shown above (Left plot) or below (Right plot) the points produced with the current technique. }
	\label{fig-bb-vs-paper}
\end{figure}

In Fig.~\ref{fig-bb-vs-paper},
\begin{figure}[htbp!]
	\centering
	\includegraphics[height=7cm]{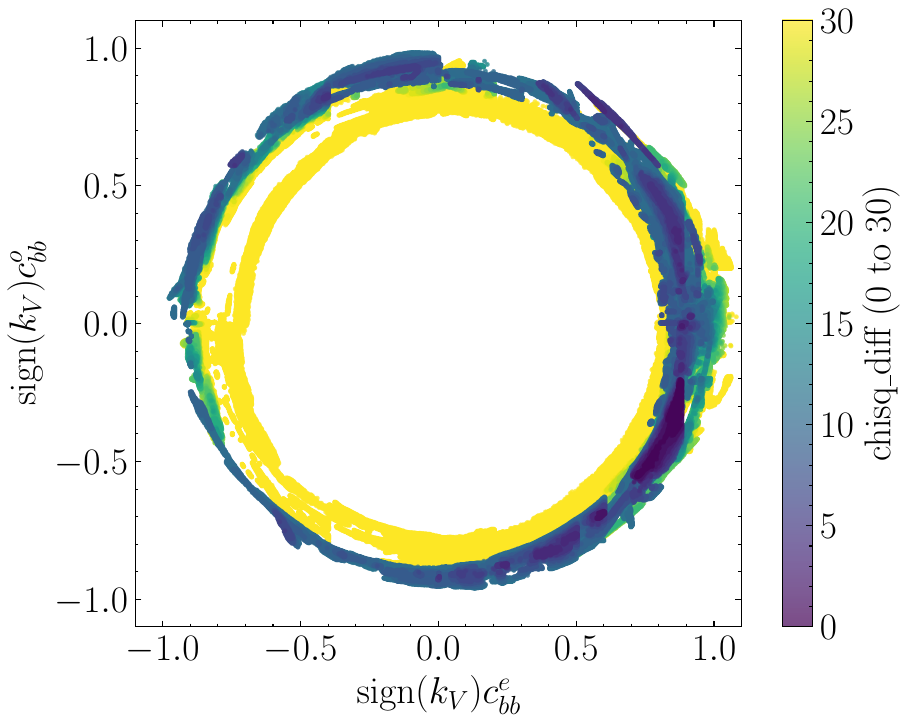} 
	\hfill
	\includegraphics[height=7cm]{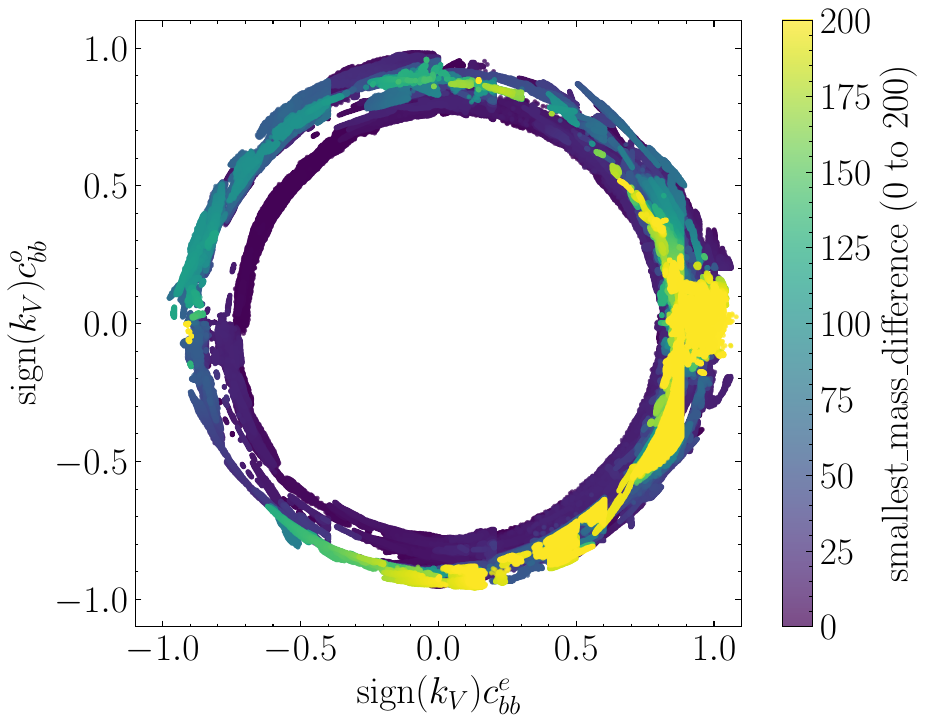}
	\caption{Combined seeded plots with CMA-ES, novelty detection and \textbf{including focus on $b\bar{b}$ couplings}. The points combine runs that required agreement with $\Delta \chi^2$ at $3\sigma$ and ones that did not. The left figure has a colour scale showing the lowest found $\Delta \chi^2$ for a given point in 2D and the right figure the highest found difference between the mass for the lightest scalar and the $125\,\textrm{GeV}$ Higgs. }
	\label{fig-bb-chisq}
\end{figure}
we present our results for the scalar and
pseudoscalar coupling of the down quarks.
The first striking difference
with respect to \cite{Boto:2024jgj}
is that we are now able to find maximal values of the pseudoscalar
couplings, even when making the stronger requirement of meeting
$\Delta \chi^2$ at $3\sigma$.
The algorithm is able to efficiently find the cancellations
needed to obey the latest eEDM limit of $4.1\times 10^{-30}\,\textrm{e.cm}$
reported by the JILA collaboration,
while simultaneously meeting every other constraint.
As shown in Ref.~\cite{Biekotter:2024ykp},
the scale at which the constants are taken matters.
For this work, we always considered the choice of the $M_Z$ scale.
We are also able to completely populate the wrong sign region. 

In broad strokes, the strategy to fill the plots was done in three main steps.
First, runs were performed with novelty reward on the parameters $|c_{bb}^o|$,
until a significant amount of converged runs that can be used for seeded runs.
This only filled the right-hand side of the plot,
with some exceptions for wrong sign with small $|c_{bb}^o|$.
Second, we added a constraint that forced $|c_{bb}^o|$ to be large,
with appropriate $C(\mathcal{O})$, and repeated the same setup of runs
with novelty reward followed by seeded runs.
Lastly, we forced $c_{bb}^e$ to have increasingly negative values, adding this
as a constraint to force the convergence in the wrong sign case
(again running seeded runs based on valid runs with novelty reward).
The last two steps each took around $10^5$ CPU hours with the first taking a fraction.
This method results in having some cuts as remnants of the fractured scans.
An improvement would be to employ novelty reward that remembers all
the previous scans, which could consist of a large amount of data. However, this significantly
increases HBOS computational time, hindering its usefulness. We leave for future
the study of alternatives. 

On the plot on the left (right) of Fig.~\ref{fig-bb-vs-paper},
the blue points from \cite{Boto:2024jgj} are drawn last (first).
We notice two features.
First,
because the (blue) points from \cite{Boto:2024jgj} were drawn starting from
the real 3HDM limit, they are concentrated in a region extending away from
the $(1,0)$ point.
And, in that way of searching, the farthest away, the more difficult it is to find a point.
The fact that we now have green points across the full circle shows the true impact of this
novel search technique.
Indeed,
by ``focusing''
(in the technical simulation sense mentioned above) on this $hbb$ plane,
we can generate allowed points on this plane with relative ease.

Second,
points with 
$\text{sgn}(k_V)c^o_{ff} > 1.0$,
found to be possible in \cite{Boto:2024jgj},
were difficult to generate here, and, by comparing green and red regions,
they require larger values of $\Delta \chi^2$.
Points with lower radius $\sqrt{|c^e_{bb}|^2 + |c^o_{bb}|^2}$
also require larger values of  $\Delta \chi^2$.
This is also apparent on the left panel of Fig.~\ref{fig-bb-chisq},
as we explain next. Our aim when starting simulations was set on enlarging the pseudoscalar component, resulting in less exploration of the region space with smaller component when comparing red and blue regions.

In Fig.~\ref{fig-bb-chisq} we use the red points from Fig.~\ref{fig-bb-vs-paper};
that is, both points that obeyed $\Delta \chi^2$ at $3\sigma$ calculated
using \texttt{HiggsSignals-2}, and points that did not.
On the left panel, we show the lowest found $\Delta \chi^2$ for a given point in 2D;
the yellow points represent $\Delta \chi^2$ equal or above $30$.
The right panel of Fig.~\ref{fig-bb-chisq} presents a colouring based on the highest
found difference between the mass for the lightest scalar and
the $125\,\textrm{GeV}$ Higgs.
We see that points with lower radius would require a second Higgs almost
degenerate with the $125\,\textrm{GeV}$ Higgs.

\subsection{$\tau$ lepton couplings}
\begin{figure}[htb]
	\centering
	\includegraphics[height=8cm]{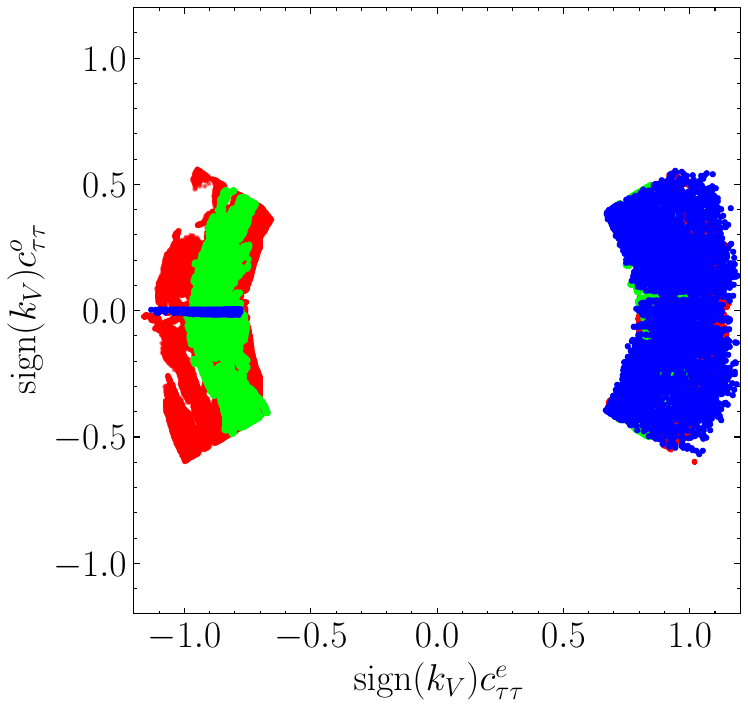} 
	\hfill
	\includegraphics[height=8cm]{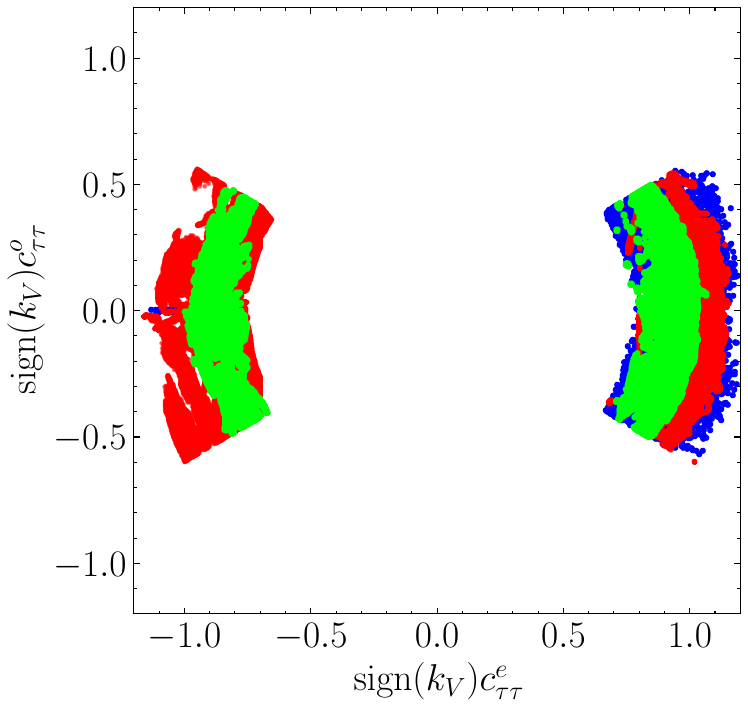}
	\caption{Combined seeded plots with CMA-ES, novelty detection and \textbf{including focus on $\tau\bar{\tau}$ couplings}. The blue coloured points from the work in \cite{Boto:2024jgj} are shown above (Left plot) or below (Right plot) the points produced with the current technique.}
	\label{fig-tautau-vs-paper}
\end{figure}

In Fig.~\ref{fig-tautau-vs-paper}, we show our results with respect to the charged leptons.
The latest data from direct searches for CP-violation in decay planes of $\tau$ leptons
places an upper limit on $c^0_{\tau\tau}/c^e_{\tau\tau} \equiv \tan{(\alpha_{h\tau\tau})}$,
yielding $|\alpha_{h\tau\tau}|<34^{\circ}$ \cite{CMS:2021sdq,ATLAS:2022akr}.
We include this as a strict constraint.
These experimental bounds are also included in \texttt{HiggsSignals-2}.

\begin{figure}[htbp!]
	\centering
	\includegraphics[height=7cm]{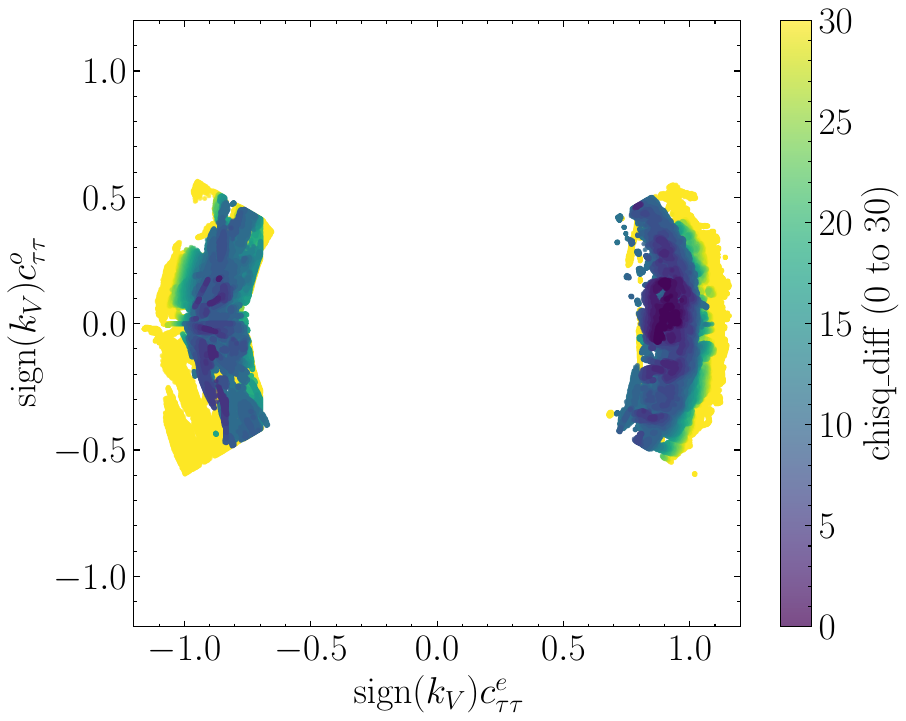}
	\hfill
	\includegraphics[height=7cm]{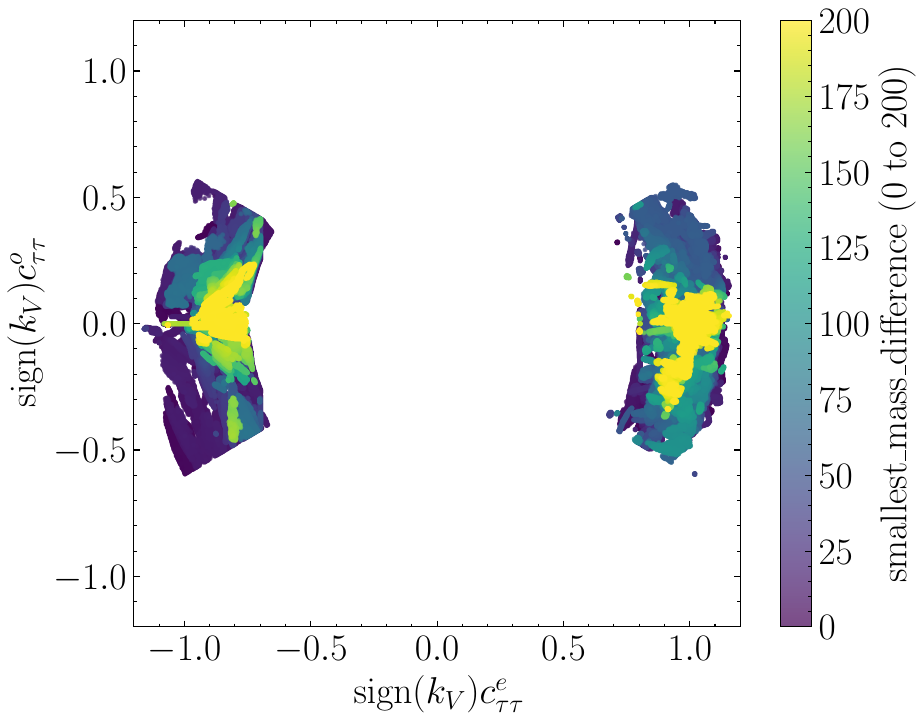}
	\caption{Combined seeded plots with CMA-ES, novelty detection and \textbf{including focus on $\tau\bar{\tau}$ couplings}. The left figure has a colour scale showing the lowest found $\Delta \chi^2$ for a given point in 2D and the right figure the highest found difference between the mass for the lightest scalar and the $125\,\textrm{GeV}$ Higgs.}
	\label{fig-tautau-chisq}
\end{figure}

The angles that meet this criteria can coexist with all other constraints.
Requiring agreement with the $\Delta \chi^2$ test at $3\sigma$ lowers the
possible radius $\sqrt{|c^e_{\tau\tau}|^2 + |c^o_{\tau\tau}|^2}$,
as also shown on the left panel of Fig.~\ref{fig-tautau-chisq}.
We are able to populate the region with negative sign for the lepton coupling,
even for large values of $|c^0_{\tau\tau}| \sim 0.5$.
As seen clearly from the line of blue points on the wrong-sign region of
Fig.~\ref{fig-tautau-chisq}-Left, this seemed completely impossible when using the
scanning method of Ref.~\cite{Boto:2024jgj}.

\subsection{Uncorrelated $\tau$ and $b$  CP-odd couplings}

\begin{figure}[htbp!]
\centering
\includegraphics[height=8cm]{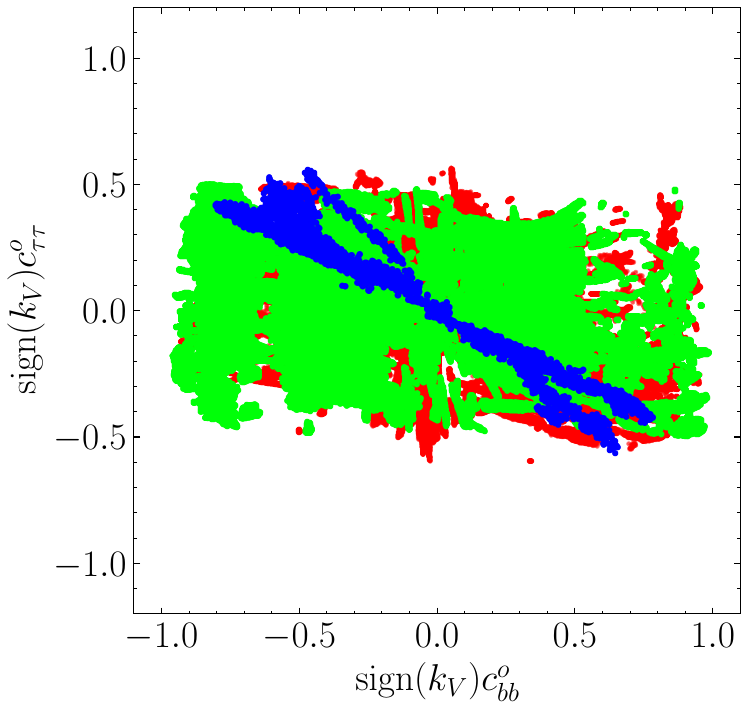}
\hfill
\includegraphics[height=8cm]{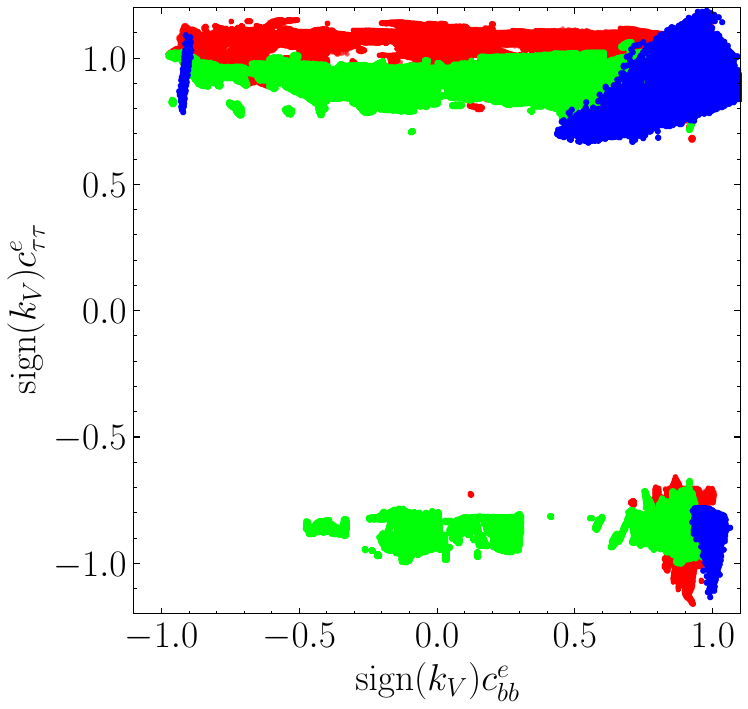} 
\caption{Combined seeded plots with CMA-ES, novelty detection
and \textbf{including focus on $\tau\bar{\tau}-b\bar{b}$ pseudoscalar couplings}.
The color code is the same as before, with blue points drawn last.}
\label{fig-bothwrong}
\end{figure}

We now turn to the first ``money-plot'' showing the full power of the
new search method, presented on the left panel of Fig.~\ref{fig-bothwrong},
where we compare $c^o_{\tau\tau}$ with $c^o_{bb}$.
This shows a very dramatic difference between the search 
performed in \cite{Boto:2024jgj}
(where one starts from points in the aligned real 3HDM and then tries to
increase $c^o_{ff}$; blue points),
and the search performed here using ML techniques (red and green points).
Indeed,
the blue points follow roughly the relation in Eq.~\eqref{notthere},
which, as pointed out in Sec.~\ref{ssub:yukawa}, would arise from taking $c^o_{tt}=0$.
This seemed to be confirmed by the simulations of \cite{Boto:2024jgj}.
Using the new method proposed in this article,
we were able to access new regions of parameter
space where Eq.~\eqref{notthere} is far from valid.
Said otherwise, we can now access regions where $c^o_{tt}$ is very different from zero
(we will come back to $c^o_{tt}$ in Sec.~\ref{ssub:top} below).
We notice from Fig.~\ref{fig-bothwrong}-Left that one can fill the whole plane,
roughly for $|c^e_{\tau\tau}|<0.5$, as limited by \cite{CMS:2021sdq,ATLAS:2022akr}, and $|c^e_{bb}|<1.0$.

Figure \ref{fig-bothwrong}-Right shows that we are able to populate the region 
that have simultaneous wrong-sign couplings for the down quarks and for the leptons,
but that, with the ranges in Eqs.~\eqref{eq:scanparameters1}-\eqref{eq:scanparameters5},
we cannot find $\textrm{sgn}(k_V)c^e_{bb} \sim -1$.

\subsection{top couplings} \label{ssub:top}

\begin{figure}[htbp!]
\centering
\includegraphics[height=8cm]{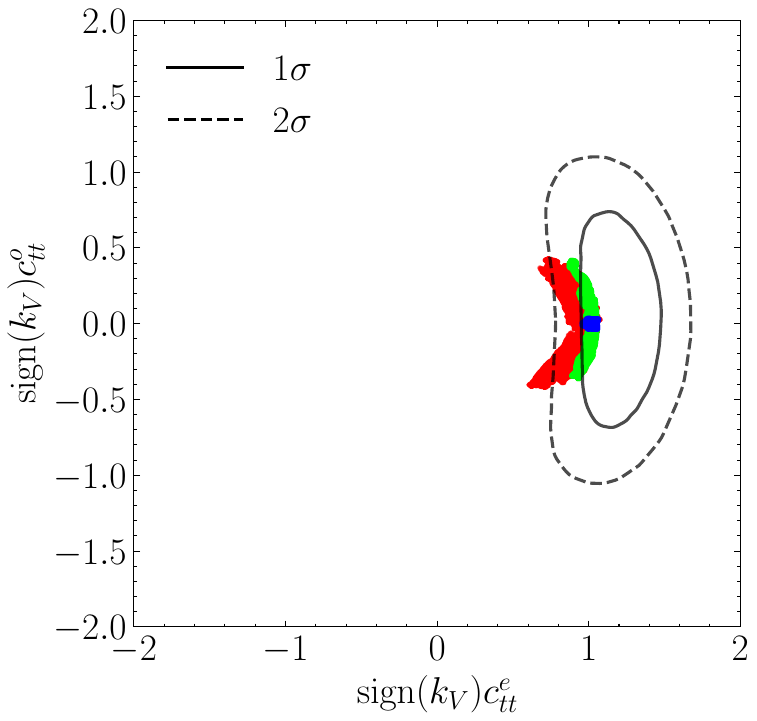}
\caption{Combined seeded plots with CMA-ES,
novelty detection and \textbf{including focus on top-top couplings}.
The color code for points is the same as before.}
\label{fig-top}
\end{figure}

In \cite{Boto:2024jgj}, the relation in Eq.~\eqref{notthere} was derived
assuming $c_{tt}^o=0$.
The scanning method used there, based on starting with points for
the real case in the alignment limit,
supported this relation and assumption.
However, our current approach finds no anticorrelation between
$c_{\tau\tau}^o$ and $c_{bb}^o$,
as shown on the left panel of Fig.~\ref{fig-bothwrong}.
This is confirmed in Fig.~\ref{fig-top},
where $|c_{tt}| > 0.3$ is a clear possibility,
when simulations cover properly the full range of parameters,
as achieved with our new method.

It turns out that the CP nature of the 125 GeV Higgs coupling to the top quark
has been probed experimentally in $tth$ production data \cite{PhysRevLett.125.061802}.
For ease of comparison,
the choice of scale in Fig.~\ref{fig-top} mimics that in \cite{PhysRevLett.125.061802}.
The experimental results are shown as $1\sigma$ (solid) and $2\sigma$ contour lines
\cite{PhysRevLett.125.061802},
indicate consistency with $c_{tt}^o \neq0$.

%
%
%

\begin{figure}[htbp!]
\centering
\includegraphics[height=8cm]{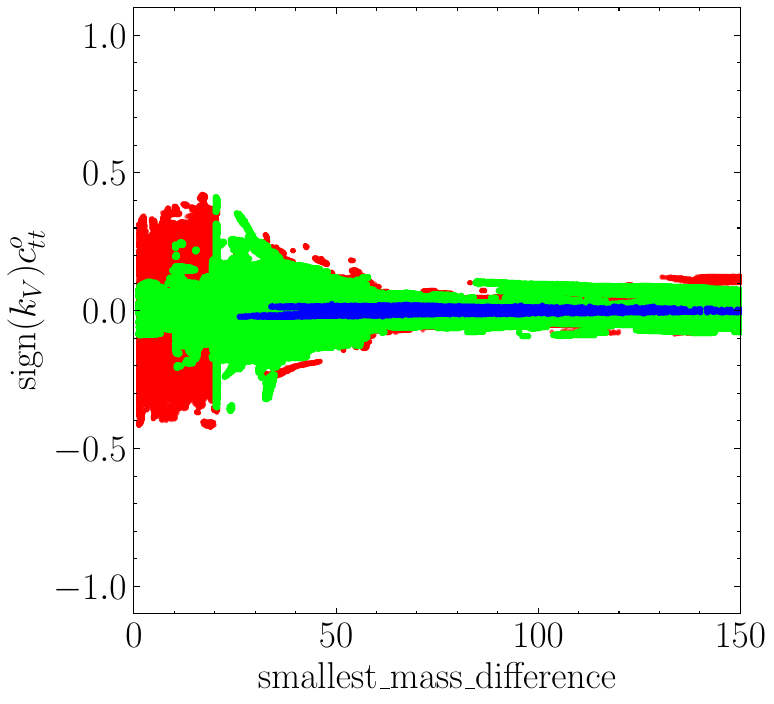}
\caption{Combined seeded plots with CMA-ES, novelty detection
and \textbf{including focus on top coupling and mass difference}.
The ``mass difference'' refers to the smallest mass difference
between every neutral scalar and $m_h=125\,\textrm{GeV}$.
The color code for points is the same as before.}
\label{fig-top-massdiff}
\end{figure}

Figure \ref{fig-top-massdiff},
shows that  increasing $c_{tt}^o$ implies the existence of a second Higgs with mass closer
to the $125 \textrm{GeV}$ Higgs.

Spurred by this plot of the C3HDM,
we have reanalyzed the C2HDM with $h_1=h_{125}$ \cite{Biekotter:2024ykp} using the
same ML scheme described in this article,
hoping that values of $|c_{tt}^o| > 0.1$ would also be possible for that simpler model.
Indeed,  
we found points where $|c_{tt}^o| > 0.1$,
but only when the lightest scalar’s mass lies below $125\,\textrm{GeV}$,
and in situations very close to degeneracy.
As pointed out in Sec.~\ref{sec:parameters},
we chose our parameter range in Eq.~\eqref{eq:scanparameters5} precisely to
exclude closely degenerate cases, thus precluding this region of the C2HDM.

\subsection{The real limit}

\begin{figure}[htbp!]
	\centering
	\includegraphics[height=8cm]{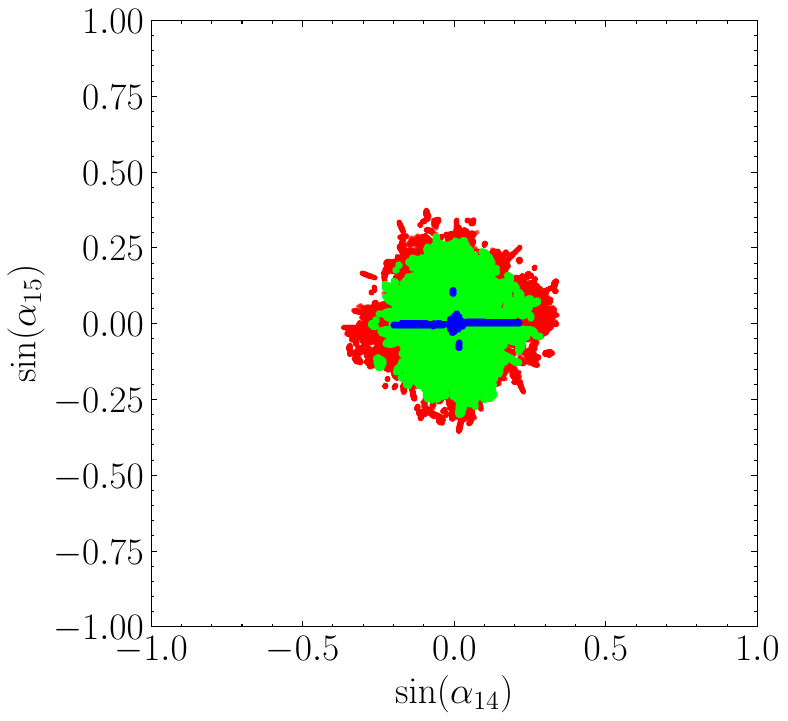}
	\hfill
	\includegraphics[height=8cm]{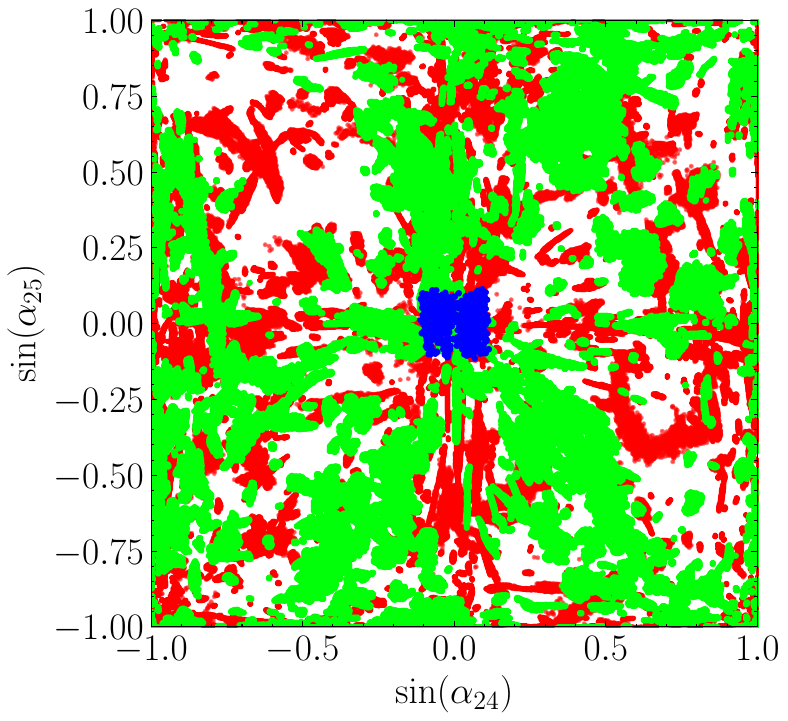}
	\caption{  Combined seeded plots with CMA-ES. The points shown \textbf{did not include dedicated scans with novelty reward focused on the parameters shown}. Blue the paper. }
	\label{fig-14-15}
\end{figure}

\begin{figure}[htbp!]
	\centering
	\includegraphics[height=8cm]{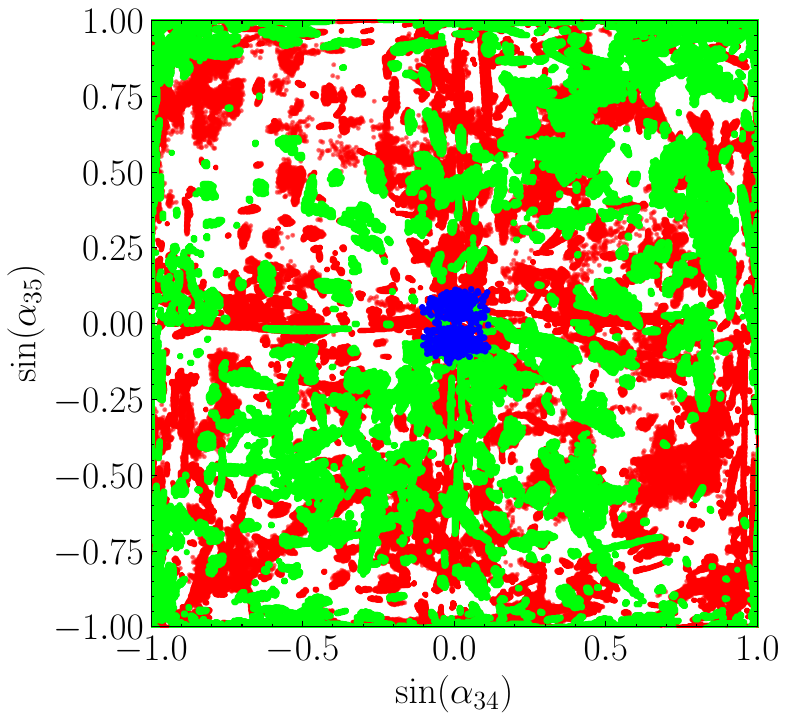}
	\hfill
	\includegraphics[height=8cm]{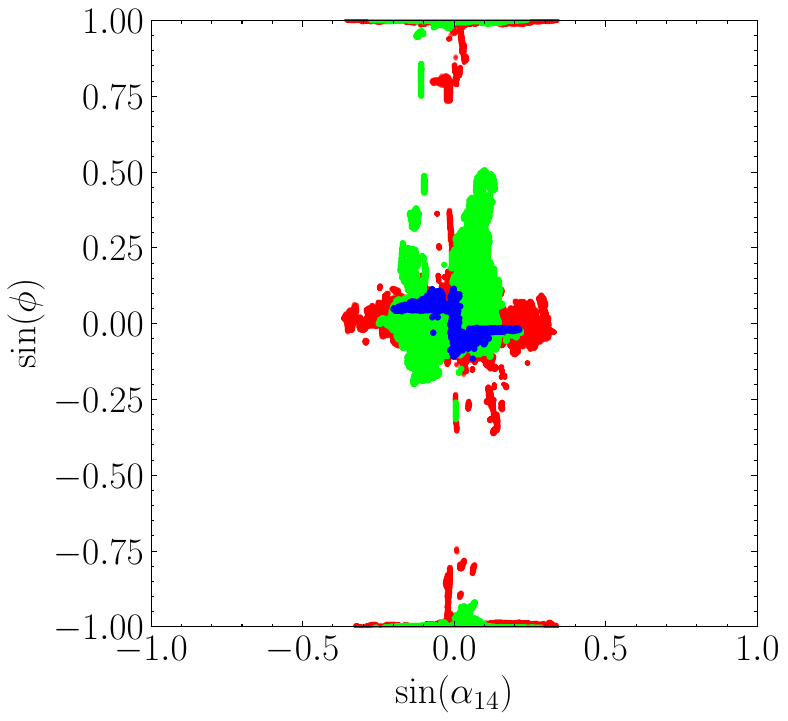}
	\caption{  Combined seeded plots with CMA-ES. The points shown \textbf{did not include dedicated scans with novelty reward focused on the parameters shown}. Blue the paper. }
	\label{fig-34-15}
\end{figure}

As introduced in Sec.~\ref{sec:physbasis},
the real limit of the model is equivalent to setting the $R_{CPV}$
as the identity matrix and $\varphi =0$.
In \cite{Boto:2024jgj},
a strategy was followed of first producing points in parameter space
for the real case,
possible when close to alignment,
then scanning around these by deviating the $R_{CPV}$ matrix in the intervals,
\begin{align}
\varphi, \alpha_{24}, \alpha_{25}, &\alpha_{34}, \alpha_{35} \in[-0.1,0.1]\, ,
\label{eq:real2}
\\*[2mm]
& \alpha_{14}, \alpha_{15} \in [-0.01,0.01]\, ,
\label{eq:real1}
\end{align}
as it was found that $\alpha_{14}, \alpha_{15}$ needed to be smaller,
in order to comply with the eEDM measurement.
In the current work, we are able to scan all the parameters in
Eqs.~\eqref{eq:real1}-\eqref{eq:real2} for the full $[-\pi, \pi]$ domain.
The set of phases corresponding to this deviation from the real limit
is plotted in Figs.~\ref{fig-14-15}-\ref{fig-34-15}.
We are able to find agreement with the statement that
$\alpha_{14}$ and $\alpha_{15}$ must be small.

Recall that
the $125\,\textrm{GeV}$ Higgs can be written as the combination
\begin{equation}
    h_1 = R_{11} x_1 +  R_{12} x_2 +  R_{13} x_3 - (R_{14} s_{\beta_1} + R_{15} c_{\beta_1} s_{\beta_2}) z_1 + (R_{14} c_{\beta_1} - R_{15} s_{\beta_1} s_{\beta_2})z_2 + R_{15} c_{\beta_2} z_3,
\end{equation}
leading to the couplings with the fermions as in Eq.~\eqref{eq:cff_odd}.
Setting $\alpha_{14}$ and $\alpha_{15}$ to be small,
reduces the (CP-odd) contributions of $z_k (k=1,2,3)$ into $h_1$,
as required. 
In contrast,
we find that the other 5 parameters can take any possible values
in the full chosen domain.
Notice the monumental difference between the blue and the red/green regions.
These are new ``money-plots'' showing the remarkable advance that the scanning strategy presented
in this article already constitutes.

Notice that the points in Figs.~\ref{fig-14-15}-\ref{fig-34-15} were obtained not by
focusing on the parameters shown on those plots (no attempt was made to cover the particular parameter
planes shown). These points were obtained by focusing on \textit{observables}, such as
$c^e_{tt}$ and $c^o_{tt}$. The fact that focusing on \textit{observables} implies automatically
very wide scanning regions for the model \textit{parameters} is a very interesting feature,
which have also found in preparatory studies using other specific models.

\section{Conclusions}\label{sec:conclusions}

We have looked at the impact of ML techniques in the full exploration of the parameter space
and physical consequences of the C3HDM.
Specifically,
we have used an Evolutionary Strategy algorithm
integrated with an anomaly detection-based Novelty Reward mechanism,
to ensure robust exploration not only of the model’s parameter space but,
crucially, of its physical implications. 

We have demonstrated the effectiveness of this new method,
by pitting it against the traditional scanning approach used in Ref.~\cite{Boto:2024jgj}.
This allowed for the efficient identification of valid points within the model parameter space.
What in Ref.~\cite{Boto:2024jgj} was a painstaking, slow search, requiring constant intervention
in the quest for new features, becomes here incomparably faster and requiring no
user supervision whatsoever.
Most importantly,
because the method converges rapidly to points with novel characteristics,
one is able to fully explore the phenomenological implications of the model.
In this respect, the differences between the blue (old method) and green (new method)
regions in Figs.~\ref{fig-bothwrong}-Left and \ref{fig-top} are impressive.

Indeed,
Fig.~\ref{fig-top} shows emphatically how the predictions of the model
obtained using the new ML methods (green and red points) far supersede the
regions obtained with a traditional scan (blue points).
Said otherwise, a traditional scan of this model might entice the idea that,
were this to be the true model of Nature, one would not find experimental
signs of large $htt$ pseudocalar couplings.
In contrast, the ML search unearths the theoretical possibility of large 
$htt$ pseudoscalar couplings in this model,
thus adding further justification for the improvement of the experimental
search for such components in $h$ production associated with $tt$
\cite{PhysRevLett.125.061802, PhysRevLett.125.061801, ATLAS:2023cbt}.

Moreover, one also finds dramatic results in the $hbb$ pseudoscalar couplings,
as seen in Fig.~\ref{fig-bb-vs-paper}-Left.
Indeed, a traditional scan - blue points in Fig.~\ref{fig-bb-vs-paper}-Left - does not yield
pure pseudoscalar $hbb$ couplings; and allows only a very small region around the wrong sign.
In stark contrast, our new ML method shows that the model is consistent with the whole
$c_{bb}^e - c_{bb}^o$ circle.
This is quite reassuring from a theoretical point of view.
Indeed, the most interesting feature of the type-Z 3HDM is that all three fermion charge sectors could,
a priori, be decoupled. In Ref.~\cite{Boto:2024jgj}, an unexpected correlation between $c_{bb}^o$
and $c_{\tau\tau}^o$ was found, and it was traced back to $c_{tt}^o \sim 0$,
as found in that article.
This corresponds to the blue points in Figs.~\ref{fig-bothwrong}-Left,
and it had the consequence that large  $c_{bb}^o$ was curbed by the experimental
bounds on large  $c_{\tau\tau}^o$ \cite{CMS:2021sdq,ATLAS:2022akr}.
The new ML method allows for large $c^o_{tt}$, thus uncorrelating 
$c_{bb}^o$ from $c_{\tau\tau}^o$, as seen on the green points in Figs.~\ref{fig-bothwrong}-Left.
Besides restoring the most interesting feature of the type-Z model,
it also urges the exploration of novel ideas to probe experimentally a large
CP-odd component in $hbb$ couplings.

We highlight that in multiple different scenarios
(large $|c^{o}_{bb}|$, large $|c^{o}_{\tau\tau}|$ and small $\Delta \chi^{2}$)
we consider additional constraints in our problem with the goal of focusing the
algorithm into regions of interest in phenomenological space.
Although the additional constraints considered inevitably reduce the
efficiency of the methodology, the fact that we are able to successfully
populate these regions without major loss in computational efficiency is
striking and a statement to the power of the methods presented here.
These finding are in agreement with similar studies with different
BSM scenarios \cite{deSouza:2022uhk,Romao:2024gjx,deSouza:2025uxb}.

Of course, the case discussed here is a particular application of the new ML method,
as it can be used to great advantage
in any model with large numbers of parameters.

The approach introduced here offers a promising direction for systematically uncovering
novel and physically meaningful solutions in complex theoretical frameworks.

\section{Acknowledgments}

We would like to thank Rui Santos and Nuno Castro for discussions.
This work is supported in part by the Portuguese
Fundação para a Ciência e Tecnologia (FCT) through the PRR (Recovery and Resilience
Plan), within the scope of the investment "RE-C06-i06 - Science Plus Capacity Building", measure "RE-C06-i06.m02 - Reinforcement of financing for International Partnerships in Science,
Technology and Innovation of the PRR", under the project with reference 2024.01362.CERN.
The work of R. Boto, J.C. Romão, and J.P. Silva is
also supported by FCT under Contracts UIDB/00777/2020, and UIDP/00777/2020.
The FCT projects are partially funded through
POCTI (FEDER), COMPETE, QREN, and the EU.
The work of M.C. Romão is also supported by the STFC under Grant No.~ST/T001011/1.
The work of R. Boto is also supported by FCT with the PhD
grant PRT/BD/152268/2021.
F.A. de Souza is also supported by FCT under the research grant with reference
No. UI/BD/153105/2022.

\bibliographystyle{JHEP}
\bibliography{ref}

\end{document}